\documentclass[a4paper,10pt]{article}
\usepackage{pos}
\usepackage{multirow}
\usepackage{pstricks}
\usepackage{psfrag}
\newcommand{\MSb}{\overline{\mathrm{MS}}}

\let\oldbibliography\thebibliography
\renewcommand{\thebibliography}[1]{\oldbibliography{#1}
\setlength{\baselineskip}{9.5pt}
\setlength{\itemsep}{2pt}} 

\title{Progress in $x$-dependent partonic distributions from lattice QCD}

\author*[a]{Krzysztof Cichy}

\affiliation[a]{Faculty of Physics, Adam Mickiewicz University, \\
ul.\ Uniwersytetu Pozna\'{n}skiego 2, 61-614 Pozna\'{n}, Poland}

\emailAdd{kcichy@amu.edu.pl}

\abstract{We review the latest progress in lattice QCD calculations of the partonic structure of hadrons. This structure is, in particular, described in terms of $x$-dependent distributions, the simplest of which are the standard parton distribution functions (PDFs). The lattice calculations rely on matrix elements probing spatial correlations between partons in a boosted hadron, that can be matched to light-cone correlations defining the relevant distributions. We discuss the recent theoretical and practical refinements of this strategy, as well as new exploratory directions. The latter include generalized parton distributions (GPDs), distributions beyond leading twist, flavor-singlet distributions and transverse-momentum dependent PDFs (TMDs). We also shortly consider the potential future impact of lattice data on phenomenology.
}

\FullConference{%
 The 38th International Symposium on Lattice Field Theory, LATTICE2021
  26th-30th July, 2021
  Zoom/Gather@Massachusetts Institute of Technology
}

\begin{document}
\maketitle

\section{Introduction}
One of the central aims of quantum chromodynamics (QCD) is to understand better structural properties of hadrons -- how their partonic constituents move and interact with one another.
Ever since the groundbreaking discovery of proton's internal structure over 50 years ago, this understanding has been progressing, but it is fair to assess that we are still rather at the beginning of the quest for full quantitative comprehension.
While we know quite a lot about the longitudinal motion of partons, our knowledge is scarce and much less quantitative when transverse degrees of freedom are in play.
This status is intimately related to the abundance of experimental data accumulated so far.
Importantly, however, recent and newly planned experiments have been developed in setups appropriate for probing the lesser known aspects of proton's structure, in particular its 3-dimensional structure.
These include the COMPASS++/AMBER experiment at CERN \cite{Adams:2018pwt}, the 12 GeV upgrade of Jefferson's Lab CEBAF accelerator \cite{Dudek:2012vr,Burkert:2018nvj}, the recently approved Electron-Ion Collider (EIC) at Brookhaven National Laboratory \cite{NAP25171,AbdulKhalek:2021gbh} and the Electron-Ion Collider in China (EicC) \cite{Anderle:2021wcy} or CERN's considered Large Hadron Electron Collider (LHeC) \cite{LHeC:2020van}.
The huge expected experimental progress should be accompanied by theoretical developments, including reliable modeling and, whenever possible, also first-principle calculations of the relevant distribution functions.
Among such functions are the well-known parton distribution functions (PDFs), characterizing the hadron only in terms of a longitudinal momentum fraction, but also functions describing the 3-dimensional structure -- generalized parton distributions (GPDs), off-forward generalization of PDFs with non-zero momentum transfer $Q^2$, and transverse-momentum dependent PDFs (TMD PDFs or TMDs), dependent on the transverse parton momentum $k_T$.
Further synthesis of GPDs and TMDs can lead to generalized TMDs (GTMDs) or, in an alternative formulation, the hadronic Wigner distribution, a mathematical 5-dimensional object summarizing quantitatively all the features of the hadron, i.e.\ its longitudinal partonic motion as well as dependence on the transverse position and momentum of partons.
All these functions are inherently non-perturbative and therefore, it is natural to expect that they can be calculated from first principles with lattice QCD (LQCD).

However, any direct extraction of partonic distribution functions from the lattice encounters the fundamental problem that these distributions are defined on the light cone, requiring Minkowski spacetime signature, opposed to the Euclidean metric required for the numerics of LQCD.
In this way, lattice investigations of nucleon structure were for a long time restricted to computations of moments of these functions, which are accessible in Euclidean spacetime.
In principle, reconstruction of the full distributions is possible from a sufficiently large number of moments, but only low ones can in practice be calculated.
The high moments suffer from very unfavorable signal-to-noise ratios and inevitable mixing with lower-dimensional power-divergent operators.
Thus, alternative approaches are needed to access the Bjorken-$x$ dependence of partonic functions.
While earlier proposals existed, a real breakthrough in $x$-dependent methods came with the seminal proposal of Ji in 2013 \cite{Ji:2013dva,Ji:2014gla}, defining so-called quasi-distributions.
They are natural analogues of partonic distributions with light-cone correlations replaced by spatial ones for a boosted hadron.
With an infinite boost, quasi-distributions become equivalent to their light-cone counterparts.
However, in an actual simulation, the boost is necessarily finite.
In such a setting, quasi-distributions are obviously different from the corresponding partonic ones, but, crucially, the two can be rigorously related.
The essential prerequisite for this is that quasi- and light-cone distributions share their infrared (IR) properties.
Then, their difference emerges only in the ultraviolet (UV) regime and is calculable perturbatively.
The possibility of such separation of scales underlies the framework of factorization that led to the viability of experimental programs involving the strong interaction.
Experimentally measured cross sections receive contributions from all energy scales or distances, but factorization separates them into short-distance and long-distance ones.
The former can be calculated in perturbation theory and the latter, the partonic distribution functions, can be treated independently.
In essentially the same manner, a quasi-distribution can be factorized into a perturbatively computable UV coefficient and the desired light-cone distribution with the same IR structure.
The analogy with factorization of experimental cross sections becomes even more tangible when one realizes that not only quasi-distributions can be related to partonic ones via factorization.
In fact, several lattice-computable observables can be subject to such factorization and enable the extraction of light-cone distributions.
For this reason, such lattice observables are sometimes called ``lattice cross sections'' (LCSs) \cite{Ma:2014jla,Ma:2017pxb}, with an additional adjective ``good'' if they satisfy some crucial properties allowing for a clean relation to their physical counterparts -- in addition to being calculable in Euclidean spacetime and having the right IR properties, they also need to have a well-defined continuum limit, i.e.\ basically they need to be renormalizable.

By now, several good LCSs have been proposed.
While Ji's seminal papers clearly led to a breakthrough, earlier proposals have also been revived in the recent years.
Historically, the first method allowing, in principle, for an access to the $x$-dependence was the hadronic tensor approach proposed by Liu and Dong in 1993 \cite{Liu:1993cv}.
It was followed by proposals to use an auxiliary quark propagator -- scalar (Aglietti et al.~\cite{Aglietti:1998ur}, 1998), heavy (Detmold and Lin \cite{Detmold:2005gg}, 2005) or light one (Braun and M\"uller \cite{Braun:2007wv}, 2007).
However, they have not initially sparked extensive practical programs aiming at a systematic extraction of partonic distributions.
The latter happened only after the 2013 proposal of Ji \cite{Ji:2013dva,Ji:2014gla} to use a Wilson line in Euclidean matrix elements, which define quasi-distributions.
Exploratory lattice studies were initiated immediately afterwards by two independent groups \cite{Lin:2014zya,Alexandrou:2015rja}, despite many key elements still missing, like renormalization and proper matching.
Since then, enormous progress has been achieved in the whole field of $x$-dependent functions.
In addition to the steady theoretical and practical progress in quasi-distributions, the approaches based on the hadronic tensor and auxiliary heavy and light quarks were reanalyzed and used for new lattice studies.
Moreover, new methods were put forward in 2017, in particular another one based on the Wilson line, the pseudo-distribution approach (Radyushkin \cite{Radyushkin:2017cyf}).
Other proposals included the ``OPE without OPE'' method based on the forward Compton amplitude (Chambers et al.~\cite{Chambers:2017dov}) and a general framework for extracting partonic distributions from several ``good LCSs'' (Ma and Qiu \cite{Ma:2014jla,Ma:2017pxb}), in particular utilizing the auxiliary light propagator, with the approach of Braun and M\"uller somewhat generalized.
It is the aim of these proceedings to review the recent efforts in extracting $x$-dependent distributions from all of the above approaches, concentrating on work of the last year.
Ealier work has been extensively reviewed in Refs.~\cite{Cichy:2018mum,Ji:2020ect,Constantinou:2020pek}.
All plots are included in this review for illustrative purposes. They come always in their arXiv versions and are reprinted under the arXiv distribution license.

\section{Brief reminder of $x$-dependent approaches}
We start with a very brief reminder of the principles of accessing $x$-dependent partonic distributions on the lattice. For details, we refer to the original publications and to the reviews~\cite{Cichy:2018mum,Ji:2020ect}.

\subsection{Generalities}
The lattice formulation is the only known way for a systematically controllable quantitative study of QCD directly from its Lagrangian.
It consists in discretizing the theory and putting its degrees of freedom on a Euclidean lattice.
In this way, the theory is regularized and amenable to numerical computations.
The Euclidean metric is necessitated by the fact the standard Minkowski one leads to highly oscillatory integrals, prohibiting any extraction of physical properties.
At the same time, the Euclidean signature is the very reason that $x$-dependent distributions cannot be directly accessed, being defined on the light cone and, thus, requiring the Minkowski metric.

However, ways for an indirect access have been found, by utilizing relations between lattice-computable observables and light-cone partonic functions.
These relations rely on the powerful framework of factorization, the same property that allows for the analysis of experimental data under the general feature of cross sections receiving contributions from a wide range of energy scales.
Naively, theoretical description of high-energy scattering processes is impossible without calculation of the non-perturbative, long-distance contributions.
Nevertheless, the key property of factorization makes it possible to separate the short-distance contributions to cross sections from the long-distance ones.
The former can be treated with perturbatively and the latter can be subject to a global fit from hundreds or thousands of experimental data, resulting in universal partonic distributions.
In a very analogous manner, lattice observables (LCSs) can be treated as experimental cross sections and factorized into perturbative and non-perturbative parts, the latter being the very same partonic functions.
This factorization procedure is usually referred to as matching.
As mentioned above, LCSs are factorizable in this way if they differ from the light-cone distributions only in the UV.
All the practically used lattice approaches to $x$-dependence satisfy this property, although the factorization may proceed at a different stage.
For example, quasi-distributions are factorized in momentum space, i.e.\ after reconstruction of the $x$-dependence, while in the pseudo-distribution approach, factorization proceeds in coordinate space, before the reconstruction.
In turn, the hadronic tensor or Compton forward amplitude can be factorized into moments of (Compton) structure functions.\footnote{Progress for these approaches is reported in the review of Ref.~\cite{Constantinou:2020pek}, including the most recent publications \cite{Liang:2019frk,Can:2020sxc}.} 

\subsection{Quasi- and pseudo-distributions}
We devote a little bit more attention to the two most popular approaches, the quasi- and pseudo-distributions.
Both of them use a non-local matrix element with the inserted quark and antiquark connected by a Wilson line to maintain gauge invariance.
For the PDF/GPD case, the matrix element has the generic form 
${\mathcal M}_{\Gamma}(z,P)=\langle N(P_3) \vert \overline{\psi}(z)\Gamma \mathcal{A}(z,0)\psi(0)\vert N(P_3)\rangle$.
$\vert N(P_3)\rangle$ is a nucleon (or other hadron) state with boost, $P_3$, conventionally taken to be in the $z$-direction. The Wilson line $\mathcal{A}(z,0)$ of length $z$ is also taken along this direction and for simplicity of notation, we only write the $z$-direction coordinates as arguments in the insertion.
The Dirac structure $\Gamma$ determines the type of accessed distribution, e.g.\ the unpolarized case is obtained with $\Gamma=\gamma_0$.\footnote{In general, different Dirac matrices can lead to the same PDFs/GPDs, but possibly with contamination from undesired distributions if a non-chiral lattice fermion is used, e.g.\ $\Gamma=\gamma_3$ leads to the unpolarized PDF/GPD, but mixing with the scalar operator needs to be taken into account \cite{Constantinou:2017sej}.}
On the lattice, the hadron needs to be created sufficiently far away from its annihilation point and from the operator insertion. 
This source-sink separation, $t_s$, determines whether one is able to isolate the contribution of the ground state without contamination from excited states.
GPDs can be obtained if one additionally switches on momentum transfer, $Q^2$, between the source and the sink.\vspace*{-3mm}

\begin{figure}[h!]
\begin{center} 
\includegraphics[scale=1]{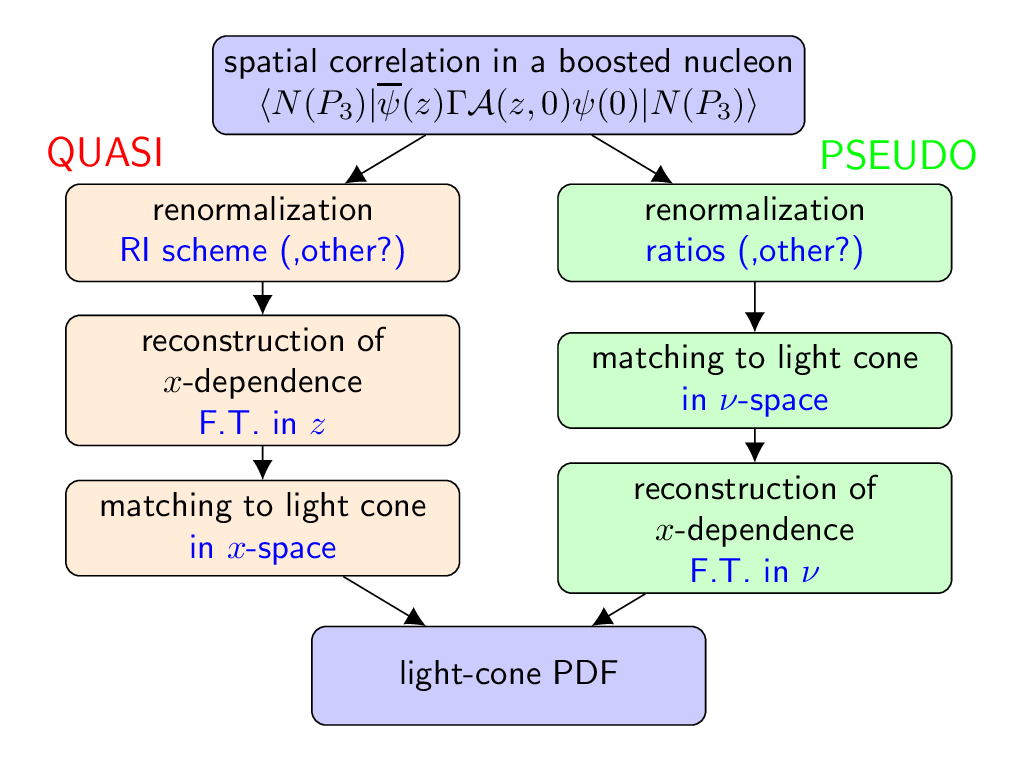}
\end{center}
\vspace*{-0.6cm}
\caption{Schematic illustration of the distribution extraction procedures for quasi- and pseudo-distributions.}
\vspace*{-3mm}
\label{fig:quasi_pseudo}
\end{figure}

We now compare the procedures that need to be followed to extract light-cone distributions from the lattice-computed bare matrix elements using both approaches, see Fig.~\ref{fig:quasi_pseudo}. 
These matrix elements contain two types of divergences, the standard logarithmic one and a power divergence induced by the Wilson line.
In principle, they can be renormalized in the same way for both approaches, but in practice the renormalization is typically done in a variant of the RI/MOM scheme for quasi-distributions, developed in Refs.~\cite{Constantinou:2017sej,Alexandrou:2017huk}, while for the pseudo-PDF case cancellation of divergences by forming a ratio is used, as suggested in Ref.~\cite{Orginos:2017kos}.
It has been recently argued that both ways can lead to non-perturbative contamination at large $z$ \cite{Ji:2020brr} and we will come back to this point below.
We emphasize, however, that the difference of the treatment of renormalization in the quasi-/pseudo-approaches is not the key difference between these methods.
This one comes at the next stage. 
In the quasi-approach, reconstruction of the $x$-dependence follows, leading to an $x$-dependent function subject to the matching procedure.
For pseudo-distributions, in turn, matching is peformed in coordinate space, at a fixed $\nu\equiv P_3z$, referred to as Ioffe time, and thus, the factorization-related objects are called Ioffe-time distributions (ITDs).
The light-cone ITDs are then finally subject to $x$-dependence reconstruction.

After all these steps, the same physical partonic distribution should be obtained in both approaches within uncertainties of the calculation.
However, these uncertainties are numerous and as of now, not fully quantified.
Their quantification is one of the crucial directions for future work and we will devote Section \ref{sec:summary} to a discussion of its prospects.
At this stage, we would like to comment on the question that has been raised in the literature and several workshops/conferences concerning $x$-dependent functions about the superiority of either of the approaches over the other one.
There is little doubt that both are valid in the sense of leading to the same light-cone distributions at large enough hadron boost.
In practice, the achievable boost is still currently relatively low on the lattice.
This means that the reconstructed and matched distributions can differ in certain $x$-regions and a comparison between such distributions coming from the same lattice data can be a very useful measure of systematics related to the procedure, in particular the factorization either in $x$-space or in $\nu$-space and usually different renormalization procedures, but also possibly other procedural differences, such as different reconstruction methods.
Thus, we can argue that the possibility of two ways of analyzing the same lattice data can and should be used to our favor, given that the calculation of the common bare matrix elements is the dominating computational cost of the whole enterprise.

\section{Dynamical progress in $x$-dependent partonic distributions}
\label{sec:results}
\vspace*{-3mm}
\begin{figure}[h!]
\begin{center} 
\hspace*{-3.5mm}
\includegraphics[scale=.53]{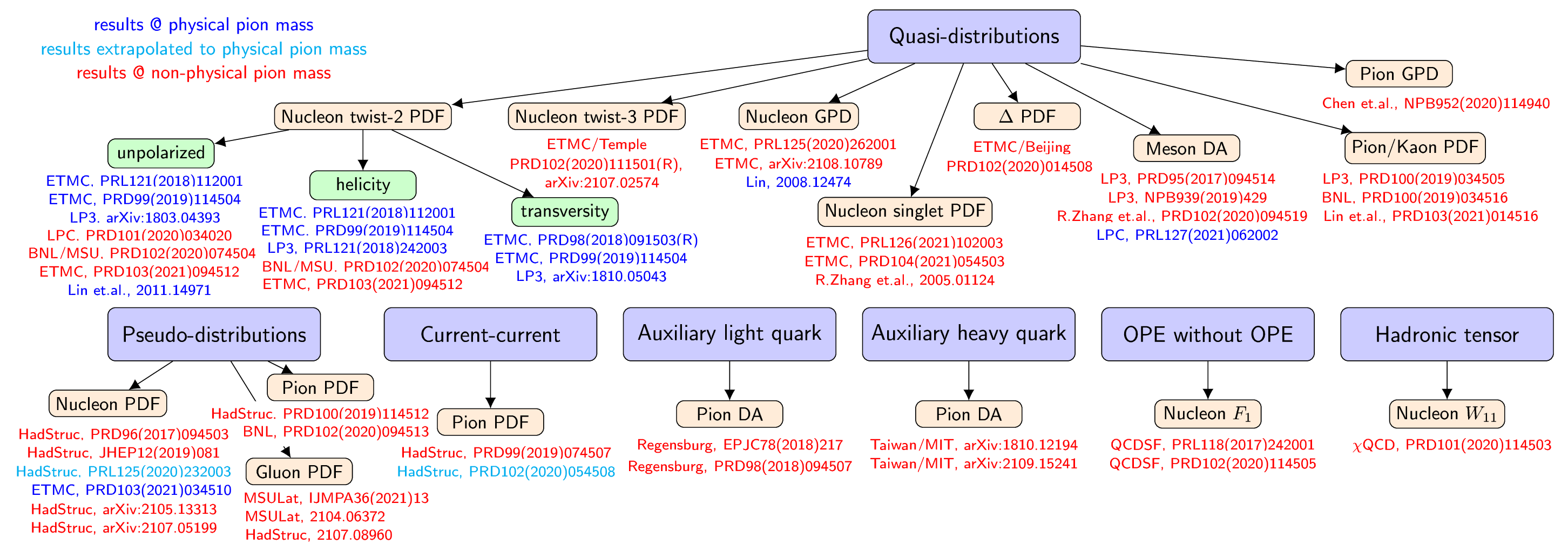}
\end{center}
\vspace*{-0.8cm}
\caption{Illustration of the dynamical progress of lattice PDFs/GPDs. Papers are grouped according to the used approach and the physical topic.}
\label{fig:summary}
\end{figure}
We now start reviewing the recent dynamical progress of lattice $x$-dependent calculations.
It is useful to realize the degree of this progress by looking at a chart itemizing papers concerning these topics from the last around three years, see Fig.~\ref{fig:summary}.
The following observations can be made.
The by far most popular approach has been the one of quasi-distributions, worked upon by the largest number of groups and encompassing a very broad set of topics.
Most work has been for the nucleon, including all leading-twist flavor non-singlet PDFs, as well as GPDs, twist-3 PDFs and flavor singlet PDFs.
Considerable amount of work has also been devoted to mesonic PDFs and distribution amplitudes (DAs), predominantly for the pion.
There is also a clear second most popular approach, pseudo-distributions, with work also by several groups and an increasing coverage of partonic functions.
All other approaches are less intensely investigated, with works only by a single group including the original authors of a given method.
Nevertheless, progress is dynamical and considerable in all approaches.
Several studies are already at an advanced stage, starting to investigate the numerous systematic effects, both practical and theoretical.
Many exploratory studies are also appearing, showing the feasibility of extraction of certain quantities despite the fact that in many cases the level of difficulty is amplified, e.g.\ by worse signal on the lattice or by additional theoretical aspects, like mixing with other distributions.

Below, we will discuss the recent highlights of the computations.
Since space for this review is limited, a subjective selection is made of topics that are covered in more detail.
Nevertheless, to keep the review as complete as possible, other recent work is also mentioned with some shorter comments.

\subsection{State-of-the-art unpolarized PDFs at the physical point}
The case of the twist-2 unpolarized PDF is the natural starting point for all approaches.
It is also somewhat special, as the case for which global fits yield the most accurate determinations.
Thus, it is the natural benchmark case for lattice determinations, with little hope of imposing constraints on this distribution, at least in the near future.
At the same time, the robustness of global fits means that any differences of the lattice-computed PDF and the phenomenological one reflects uncontrolled uncertainties in the former.
In this way, it helps to assess the different types of these uncertainties and at some point, it can lead to establishing the appropriate and robust setup for the lattice calculations with respect to several choices that need to be made on the lattice, in terms of lattice parameters, but also more fundamental procedural ingredients, like the approach to non-perturbative renormalization or the order of perturbative corrections in the matching.\vspace*{-2mm}

\begin{figure}[h!]
\begin{center} 
\hspace*{5.2cm}
\includegraphics[scale=0.51216, angle=0]{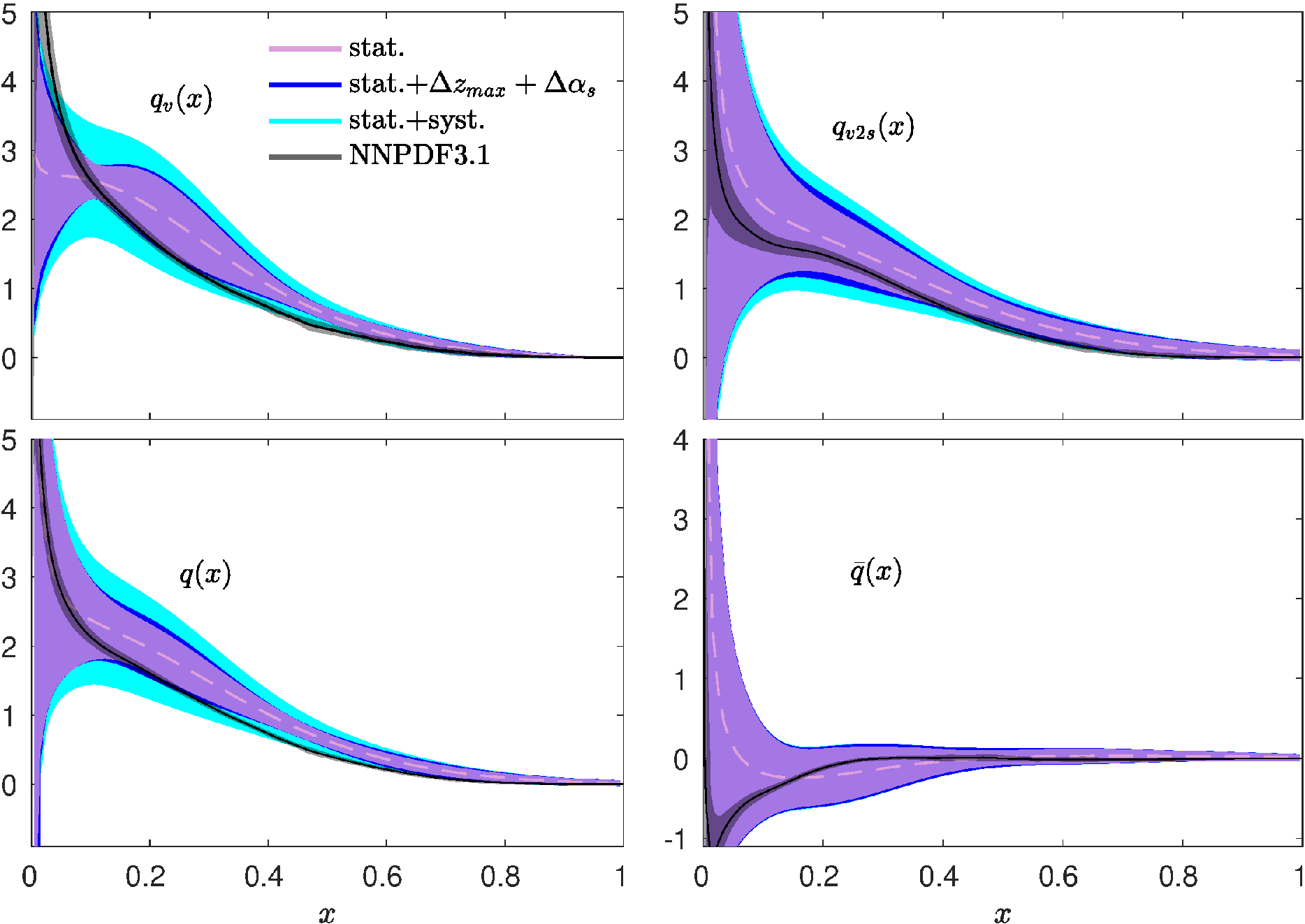}
\setlength{\tabcolsep}{3pt}
\rput(-5.2cm,5.9cm){\includegraphics[height=3.2277cm,width=5.527cm,angle=0]{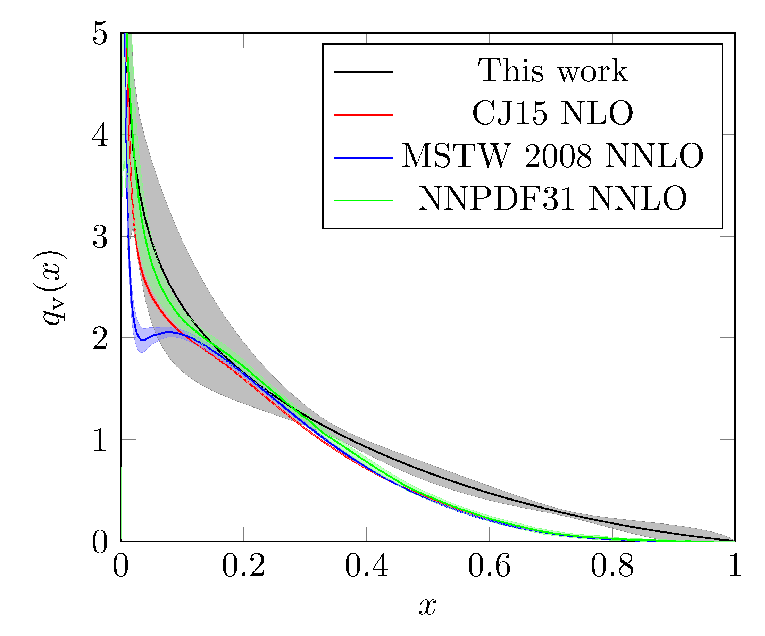}}
\rput(-5cm,3.7cm){\scriptsize\begin{tabular}{|c|c|c|c|}\hline $\uparrow$ & \cite{Joo:2020spy} & PSEUDO & clover \\\hline \multicolumn{3}{|c|}{$m_\pi\!=\!358,278,172$ MeV} & $a\!=\!0.094$ fm\\\hline\end{tabular}}
\rput(-4.7cm,2.6cm){\scriptsize\begin{tabular}{|c|c|c|c|}\hline $\rightarrow$ & \cite{Bhat:2020ktg} & PSEUDO & TMF \\\hline \multicolumn{3}{|c|}{$m_\pi\!=\!130$ MeV} & $a\!=\!0.094$ fm\\\hline\end{tabular}}
\end{center}
\vspace*{-1cm}
\caption{State-of-the-art unpolarized isovector PDFs of the nucleon at or extrapolated to the physical pion mass, from the pseudo-PDF approach. Left: the valence distribution $q_v$ (Jo\'o et al.~\cite{Joo:2020spy}), right: $q_v$ (upper left), $q_{v2s}\equiv q_v+2\bar{q}$ (upper right), $q\equiv q_v+\bar{q}$ (lower left), $\bar{q}$ (lower right) (Bhat et al.~\cite{Bhat:2020ktg}). The vertical axes of both studies for $q_v$ are aligned to allow for a better comparison.}
\vspace*{-3mm}
\label{fig:unpol}
\end{figure}

We begin by reviewing the two recent advanced studies of the unpolarized isovector PDF obtained using the pseudo-distribution approach.
Given the special status of the unpolarized PDF argued above, these results allow us to draw very general conclusions about the current stage of $x$-dependent distributions from the lattice, valid also for several other computations reported below.

Jo\'o et al.~\cite{Joo:2020spy} used clover fermions with three values of the pion mass ranging from 358 down to 172 MeV. 
They found the pion mass effects to be rather small and extrapolated the final results to the physical point, testing fitting ansatzes linear and quadratic in the pion mass. 
The $x$-dependence was reconstructed with a fitting ansatz of the form $q_v\sim x^a(1-x)^b$ with polynomial corrections in $\sqrt{x}$.
The resulting valence distribution, $q_v$, is shown on the left of Fig.~\ref{fig:unpol}.
Compared to phenomenological PDFs, there is rather good agreement at $x\lesssim0.4$, with the lattice PDF significantly too high for larger $x$.
The plotted errors are only statistical and thus, it is clear the disagreement ensues from unquantified systematics, such as discretization effects or 1-loop truncation of the matching.
A different approach to systematic effects was adopted by Bhat et al.~\cite{Bhat:2020ktg}.
They used the lattice data of a previous ETMC lattice study in the quasi-distribution framework \cite{Alexandrou:2018pbm,Alexandrou:2019lfo}, using twisted mass fermions (TMF) directly at the physical point.
They also adopted a fitting ansatz reconstruction (without polynomial corrections in $\sqrt{x}$, found to be statistically insignificant) and within statistical uncertainties obtained a very similar valence distribution to the one of Jo\'o et al., see the upper left plot of the right part of Fig.~\ref{fig:unpol}.
Here, however, the disagreement with global fits was found to be somewhat smaller and only in the intermediate-$x$ region (purple band).
Nevertheless, Bhat et al.\ attempted to estimate the unquantified systematics with ``plausible'' scenarios for systematic effects, as proposed in Ref.~\cite{Cichy:2019ebf}.
This consisted in assuming some magnitude for cutoff, finite volume, excited states and truncation effects, with this magnitude taken at the typical level of other hadron structure studies.
Some systematics could also be quantitatively estimated.
Within such combined statistical and ``plausible'' systematic uncertainty (cyan band), $q_v$ was found to agree with global fits for the whole $x$-range.
Similar conclusions were drawn for the other considered non-singlet distributions, involving also antiquarks, with $\bar{q}$ and $q_v+2\bar{q}$ already agreeing with global fits within the statistical, but larger than for $q_v$, errors.

As mentioned above, these calculations represent the typical conclusions drawn from present-day lattice studies.
Predominantly, this concerns the fact of their qualitative agreement with phenomenology, with some quantitative tension at least in some regimes.
Additionally, it is not yet clear which systematic effects are most important and thus, their careful investigation is mandatory before attempting to draw final quantitative conclusions.
However, the clearly positive message is that typical sizes of systematic effects, as seen in various lattice computations, can yield full agreement with phenomenology, i.e.\ one does not have reasons to expect particularly large size of these effects.
Nevertheless, the ``plausible systematics'' approach can not replace a proper study of several sources of systematics. 
An example of such a source, obvious for lattice practicioners, are discretization effects, to be discussed below.

\subsection{Lattice data and NNPDF/JAM reconstruction of distributions}
\begin{figure}[h!]
\begin{center} 
\includegraphics[scale=0.46, angle=0]{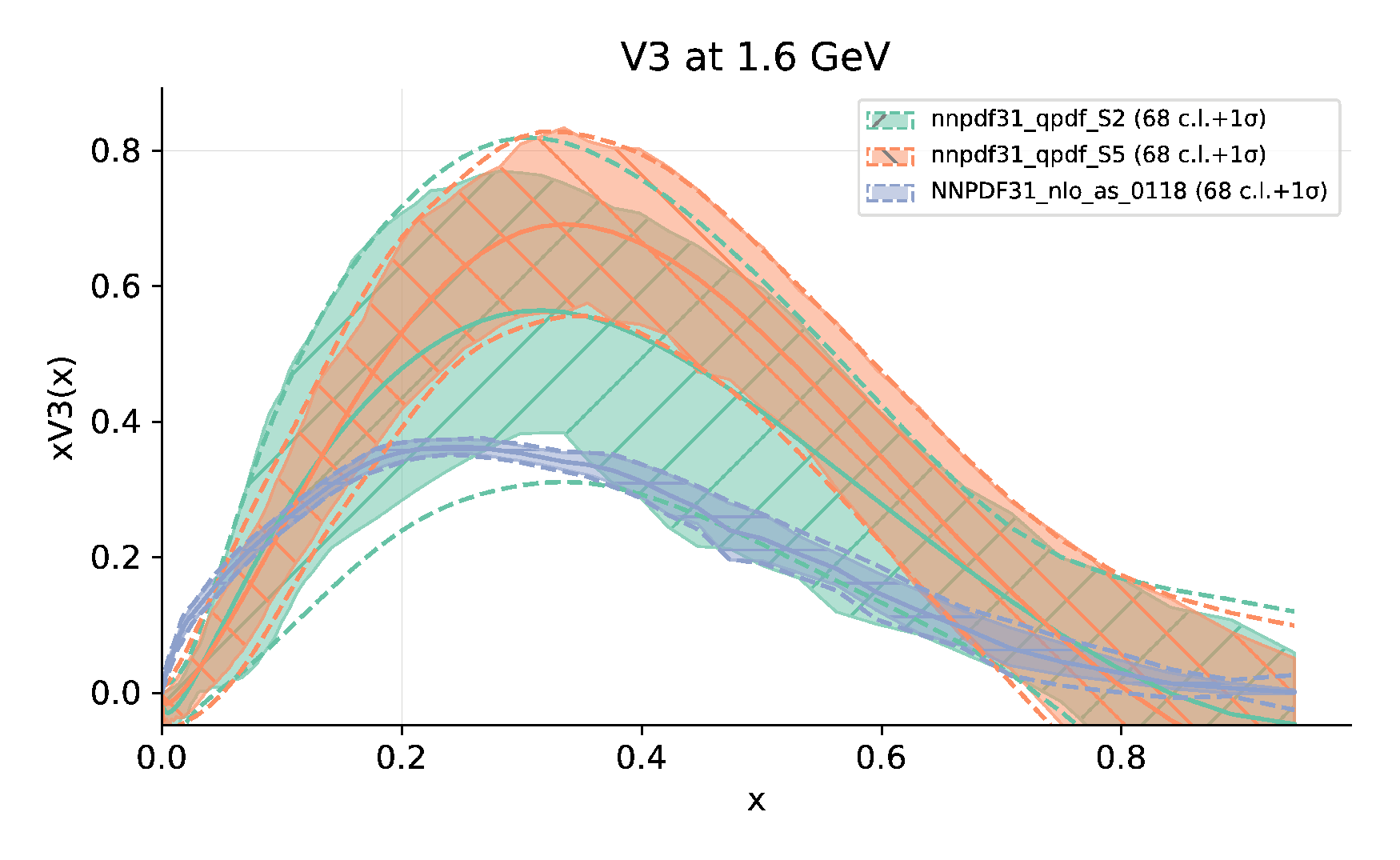}
\hspace*{-6mm}
\includegraphics[scale=0.46, angle=0]{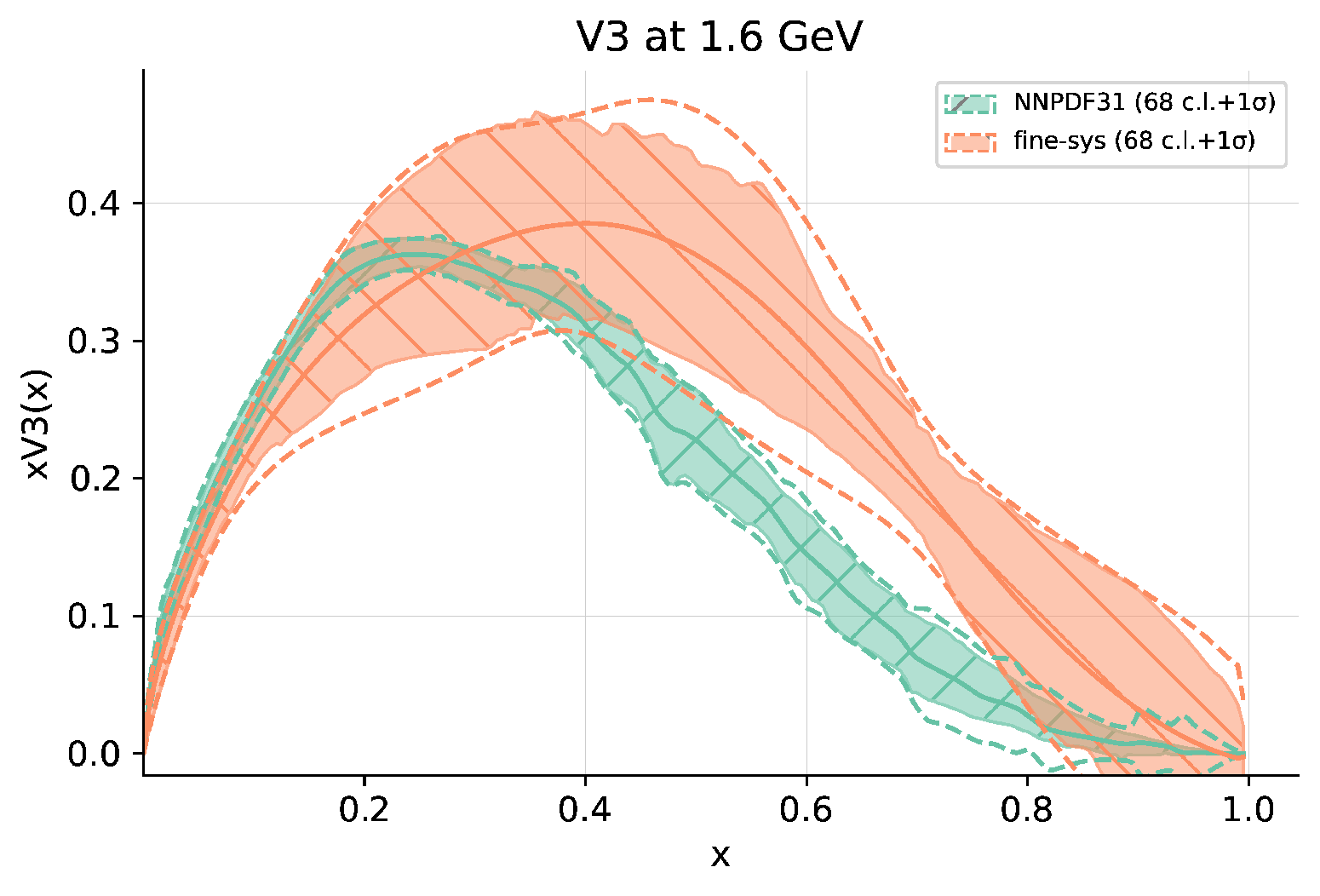}
\setlength{\tabcolsep}{3pt}
\rput(4cm,5.9cm){\scriptsize\begin{tabular}{|c|c|c|c|}\hline $\downarrow$ & \cite{DelDebbio:2020rgv} & PSEUDO & clover \\\hline \multicolumn{3}{|c|}{$m_\pi\!=\!358,278,172$ MeV} & $a\!=\!0.094$ fm\\\hline\end{tabular}}
\rput(-3.4cm,5.9cm){\scriptsize\begin{tabular}{|c|c|c|c|}\hline $\downarrow$ & \cite{Cichy:2019ebf} & QUASI & TMF \\\hline \multicolumn{3}{|c|}{$m_\pi\!=\!130$ MeV} & $a\!=\!0.094$ fm\\\hline\end{tabular}}
\end{center}
\vspace*{-1cm}
\caption{Unpolarized isovector valence PDF $V_3$ of the nucleon, reconstructed via the NNPDF framework. Left: from Ref.~\cite{Cichy:2019ebf}, using ETMC lattice data of Refs.~\cite{Alexandrou:2018pbm,Alexandrou:2019lfo} and the quasi-PDF factorization + ``plausible systematics'' for cutoff effects, FVE, excited states and truncation effects. Right: from Ref.~\cite{DelDebbio:2020rgv}, using HadStruc lattice data \cite{Joo:2020spy} and the pseudo-PDF factorization + estimates of systematics from comparing two lattice spacings, two volumes and two pion masses.}
\vspace*{-3mm}
\label{fig:unpol2}
\end{figure}

The lattice data related to the lattice studies of the previous subsection were also analyzed in the framework of the robust NNPDF and JAM global fitting methodologies.
The first of these investigations was carried out in Ref.~\cite{Cichy:2019ebf} using the matrix elements calculated in Refs.~\cite{Alexandrou:2018pbm,Alexandrou:2019lfo}.
The authors started with closure tests, consisting in producing mock data for quasi-PDF matrix elements from a variant of NNPDF3.1 unpolarized PDFs \cite{Ball:2017nwa}.
The procedure encompassed DGLAP evolution from 1.65 to 2 GeV, inverse 1-loop matching and inverse Fourier transform, for 16 values of $z$ (16 real + 15 imaginary).
Then, the matrix elements were used in a reverse procedure to reconstruct light-cone PDFs, involving a Fourier transform via a neural-network fit, 1-loop matching and again DGLAP evolution, from 2 to 1.65 GeV.
The reconstructed PDFs were found to be in agreement with the initial NNPDF.
Despite the fact that such an exercise may seem trivial, it actually shows the power of the convolution involved in relating the quasi- and light-cone distributions, since the information contained in the 16 complex matrix elements was enough to get the physical PDF, as contrasted with thousands of experimental measurements typically involved in global fits.
Obviously, the current precision of lattice data (statistical and, in particular, systematic) is much worse than the one of the mock data of this exercise.
Hence, the actual lattice data were also put through the NNPDF machinery.
As expected, tensions were revealed in a wide range of $x$-values.
Using the ``plausible systematics'' method, originally introduced in this work, quantitative agreement could be reached, but only within very large uncertainties, see Fig.~\ref{fig:unpol2} (left).

\hspace*{6mm}

The NNPDF framework was also used \cite{DelDebbio:2020rgv} for the HadStruc collaboration data of Ref.~\cite{Joo:2020spy}.
This allowed to test the combination with another convolution relating Euclidean and light-cone distributions, the one of the pseudo-PDF method, entering in $\nu$-space of ITDs.
In this case, the authors opted for a partial investigation of systematics, including cutoff effects from comparing two lattice spacings (0.094 and 0.127 fm), pion mass effects from comparing $m_\pi=172$ and $278$ MeV results and FVE from two volumes, 3 and 4 fm (at $m_\pi=415$ MeV).
This enlarged the statistical uncertainties, but was not enough to find full quantitative agreement with the global NNPDF fit, see Fig.~\ref{fig:unpol2} (right).
It is clear that such a procedure is more in spirit of typical lattice studies than the ``plausible systematics'' one, but it is unsurprising that the only partial inclusion of systematics, e.g.\ only two rather coarse lattice spacings, is not enough to account for all tension between the current lattice results and global fits.

\begin{figure}[h!]
\begin{center} 
\includegraphics[scale=0.218, angle=0]{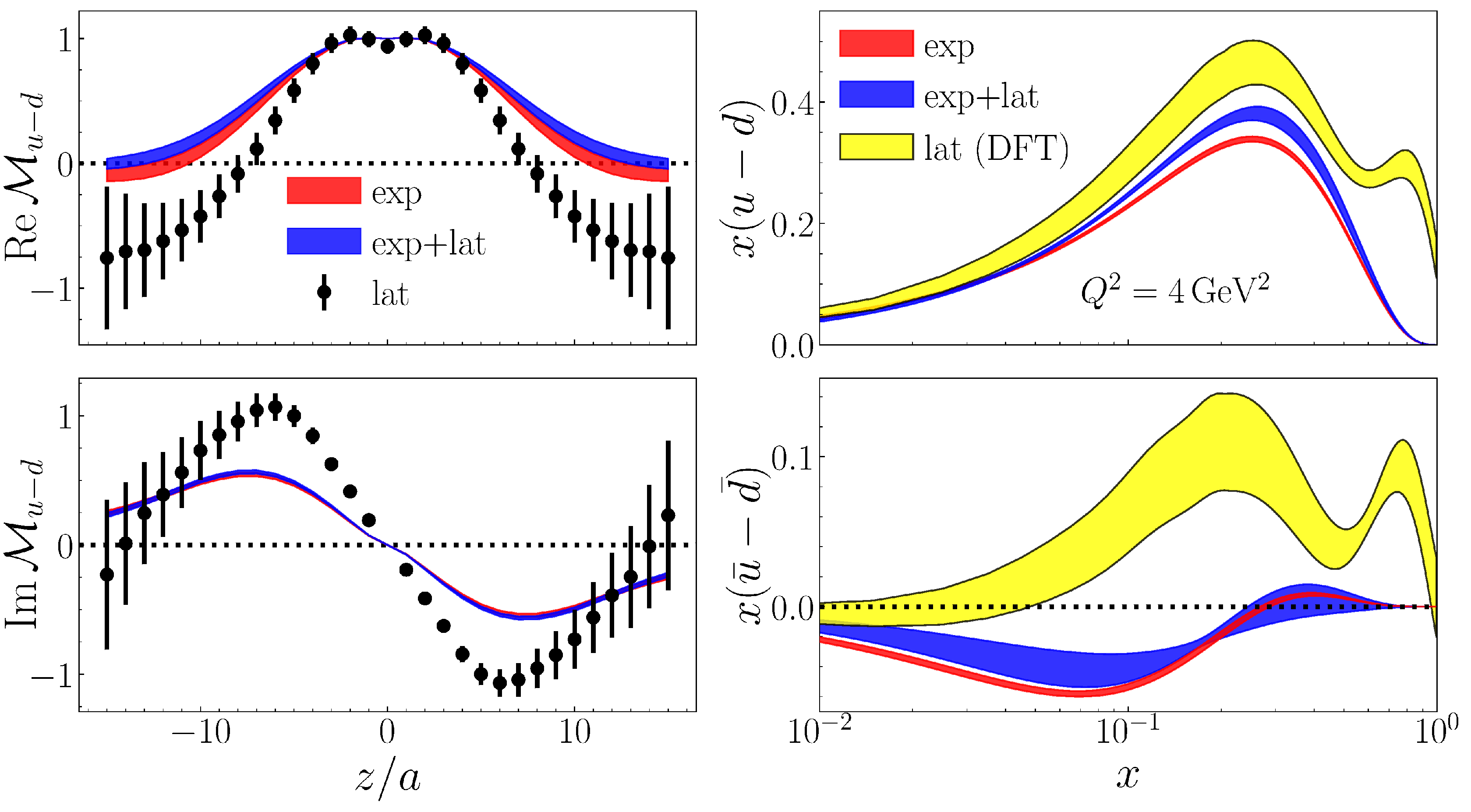}
\hspace*{-2mm}
\includegraphics[scale=0.218, angle=0]{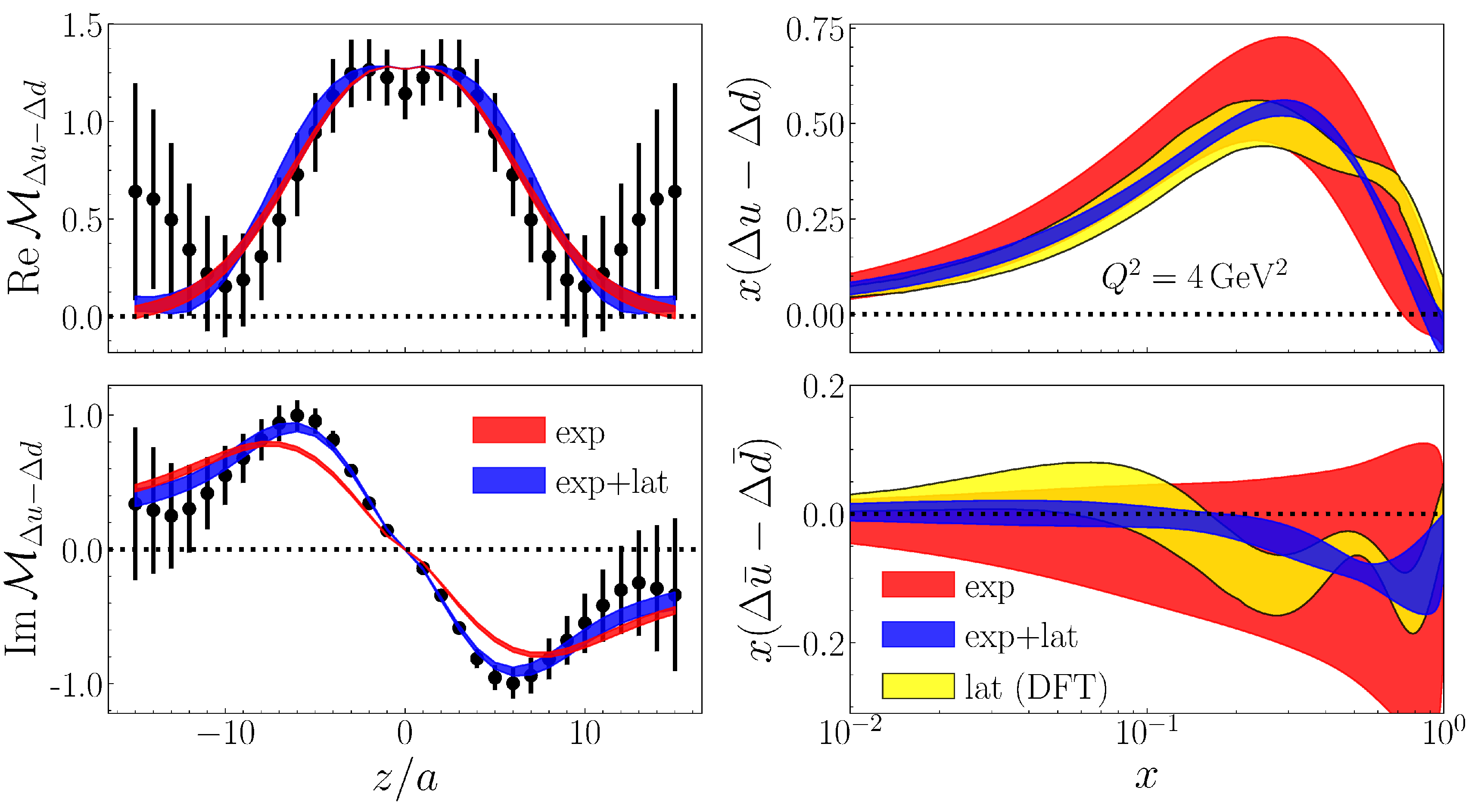}
\setlength{\tabcolsep}{3pt}
\rput(0cm,4.76cm){\scriptsize\begin{tabular}{|c|c|c|c|c|c|}\hline $\downarrow$ & \cite{Bringewatt:2020ixn} & QUASI & TMF & $m_\pi\!=\!130$ MeV & $a\!=\!0.094$ fm\\\hline\end{tabular}}
\end{center}
\vspace*{-1cm}
\caption{Unpolarized and helicity isovector PDFs, reconstructed with the JAM framework \cite{Bringewatt:2020ixn}, using ETMC lattice data of Refs.~\cite{Alexandrou:2018pbm,Alexandrou:2019lfo}. The first/third column shows the real (top) and imaginary (bottom) part of renormalized unpolarized/helicity matrix elements. The second/fourth column shows the reconstructed unpolarized/helicity distributions for quarks (top) and antiquarks (bottom). Different bands/datapoints represent only experimental (red), only lattice (black or yellow) or combined experimental+lattice (blue) data used in the JAM reconstruction.}
\vspace*{-3mm}
\label{fig:JAM}
\end{figure}

The same ETMC data were also used in within the JAM framework \cite{Bringewatt:2020ixn}, considering in addition the helicity PDFs case.
The lattice data of Refs.~\cite{Alexandrou:2018pbm,Alexandrou:2019lfo} were for the first time actually treated on the same footing as experimental DIS cross sections and their constraining power was evaluated, see Fig.~\ref{fig:JAM}.
In the unpolarized case, there is again significant tension between lattice and experiment, obvious already at the level of matrix elements.
The lattice data is concluded to need greatly improved precision to have any influence on the global fit.
In turn, for the helicity case, promising agreement between lattice and experiment is established, with the current lattice precision already providing constraints.
Thus, one can foresee that lattice computations with fully controlled systematics can exert some influence on global helicity PDFs.

The final conclusion from these NNPDF/JAM reconstructions is that the unpolarized PDFs case should most appropriately be treated as a benchmark case, with the lattice trying to establish agreement with the phenomenological PDFs, by improving precision and controlling systematic effects, but without attempting to provide quantitative constraints.
Establishing such an agreement would demonstrate the possibility for quantitative control over the lattice results and this would allow to argue for the plausibility of drawing quantitative conclusions for the cases of partonic functions that are much less known from global fits to experimental data.
The latter include already helicity PDFs, but in particular transversity PDFs, so far very weakly constrained by experiment, and other distributions with a similar or even more volatile status, like higher-twist PDFs, GPDs or TMDs.

\subsection{Role of discretization effects}
Until recently, most lattice studies involved simulations at only one value of the lattice spacing.
Thus, the final results were necessarily contaminated with discretization effects of unknown magnitude, possibly $x$-dependent.
Hence, it is obvious and pressing to evaluate this magnitude and perform a continuum limit extrapolation, something that was indeed attempted by some groups.

Before we review this work, it is important to point out that the non-local operators occuring in the relevant matrix elements induce $\mathcal{O}(a)$ cutoff effects. 
Ref.~\cite{Green:2020xco} analyzed this issue in detail, using an auxiliary field approach.
It was concluded that $\mathcal{O}(a)$ terms appear even in a chirally-symmetric lattice setup and also, that the automatic $\mathcal{O}(a)$-improvement of TMF does not prevent these, although it may eliminate some of the contributions and reduce the number of necessary improvement coefficients.
The paper also outlined a general framework for achieving the $\mathcal{O}(a)$-improvement, with the possibilities of obtaining the improvement coefficients from chiral Ward identities, lattice perturbation theory and/or numerically.

\begin{figure}[h!]
\begin{center} 
\includegraphics[scale=0.3, angle=0]{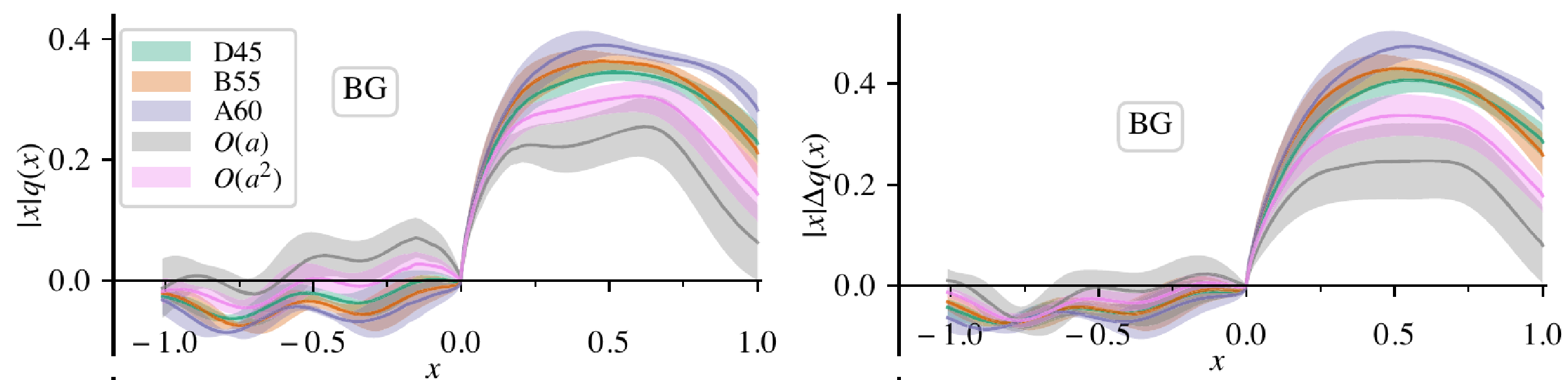}
\hspace*{-2mm}
\includegraphics[scale=0.3, angle=0]{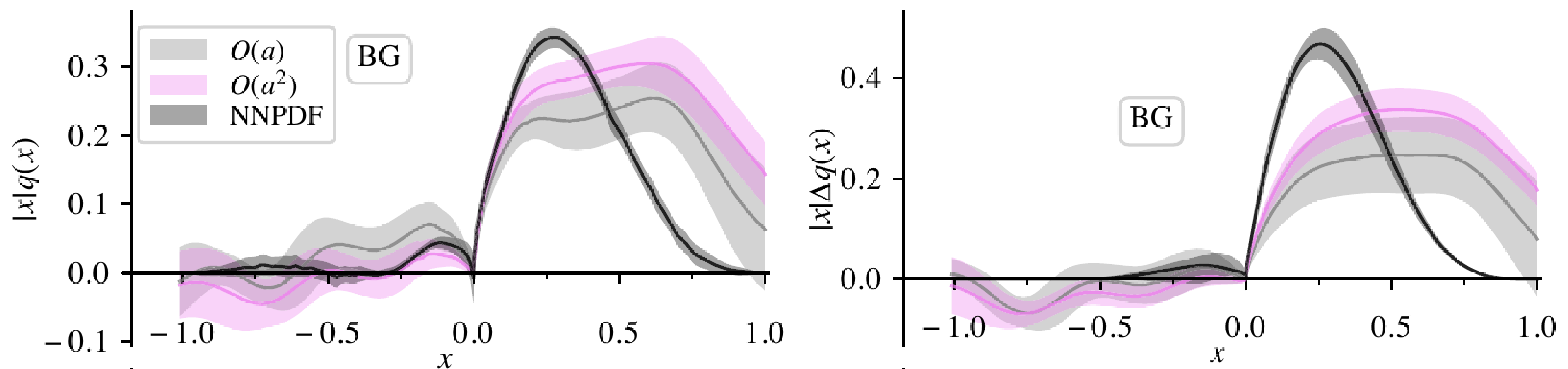}
\setlength{\tabcolsep}{3pt}
\rput(-6.2cm,6.25cm){\scriptsize\begin{tabular}{|c|c|c|c|c|c|}\hline $\downarrow$ & \cite{Alexandrou:2020qtt} & QUASI & TMF & $m_\pi\!=\!370$ MeV & $a\!=\!0.064,0.082,0.093$ fm\\\hline\end{tabular}}
\end{center}
\vspace*{-5mm}
\caption{ETMC study of discretization effects in quasi-PDFs \cite{Alexandrou:2020qtt}. Unpolarized (left) and helicity (right) isovector PDFs. The upper row shows results for the 3 considered ensembles, together with their $\mathcal{O}(a)$ or $\mathcal{O}(a^2)$ continuum extrapolation. The lower row compares the continuum distributions with NNPDF global fits.}
\vspace*{-3mm}
\label{fig:cont1}
\end{figure}

Meanwhile, the existing continuum limit studies are bound to assume the presence of $\mathcal{O}(a)$ effects.
A recent example is the ETMC twisted-mass study \cite{Alexandrou:2020qtt} of unpolarized and helicity PDFs at a non-physical pion mass of around 370 MeV and three lattice spacings from 0.064 fm to 0.093 fm, see Fig.~\ref{fig:cont1}.
The general feature of the results is a quite significant dependence on the lattice spacing and a tendency for better agreement with phenomenological fits after elimination of cutoff effects.
In particular, the latter were shown to be able to change the sign of the antiquark distribution, explaining the apparently wrong sign obtained in earlier ETMC investigations \cite{Alexandrou:2018pbm,Alexandrou:2019lfo}.
Nevertheless, the continuum-extrapolated distributions are significantly off from the global-fits PDFs, indicating again that no single source of systematics may explain the discrepancy.

\begin{figure}[h!]
\begin{center} 
\includegraphics[scale=0.33, angle=0]{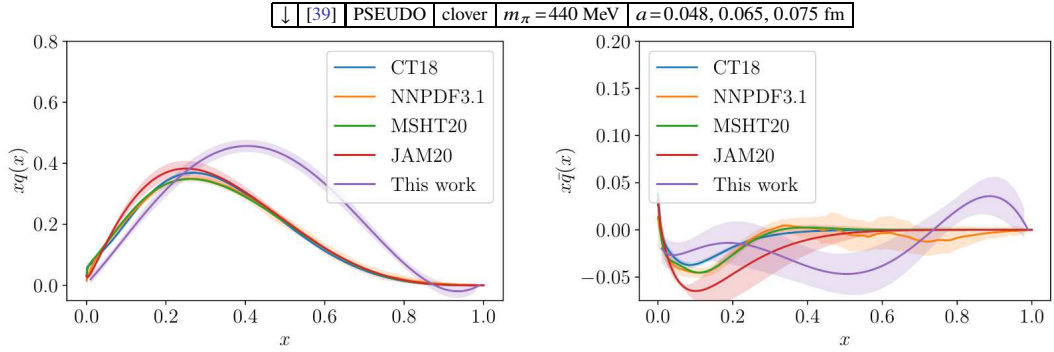}
\setlength{\tabcolsep}{3pt}
\rput(-6.7cm,4.5cm){\scriptsize\begin{tabular}{|c|c|c|c|c|c|}\hline $\downarrow$ & \cite{Karpie:2021pap} & PSEUDO & clover & $m_\pi\!=\!440$ MeV & $a\!=\!0.048,0.065,0.075$ fm\\\hline\end{tabular}}
\end{center}
\vspace*{-7mm}
\caption{HadStruc study of discretization effects in pseudo-PDFs \cite{Karpie:2021pap}. The left/right panel shows the quark/antiquark unpolarized isovector PDF in the continuum limit, as compared to various phenomenological fits. }
\vspace*{-3mm}
\label{fig:cont2}
\end{figure}

The first continuum-limit study of the unpolarized isovector PDF using the pseudo-distribution approach appeared very recently, using three ensembles of clover fermions with lattice spacings between 0.048 fm and 0.075 fm, at a pion mass of 440 MeV.
The authors used the summed generalized eigenvalue problem (sGEVP) for a good control over excited states and introduced a novel technique for parametrizing systematic errors employing Jacobi polynomials.
Using a Bayesian approach, they analyzed scenarios for including different combinations of $\mathcal{O}(a/|z|)$ and $\mathcal{O}(a\Lambda_{\rm QCD})$ cutoff effects, as well as $\mathcal{O}(z^2\Lambda_{\rm QCD}^2$ higher-twist effects (HTE).
The first type of cutoff effects was found to be significant, while the fits were basically insensitive to the other type of discretization effects and to HTE.
It is, however, clear that this is a matter of statistical precision of data, e.g.\ HTE are expected to be the largest in the regime of $z$ that currently has large errors.
The final continuum-extrapolated PDFs are shown in Fig.~\ref{fig:cont2} and compared to phenomenological extractions, showing a similar overall picture as in the ETMC study.
Note that the discrepancy is particularly visible in the large-$x$ region and, thus, must at least partially come from the non-physically heavy pion mass, which is known to enlarge the values of PDFs moments.\vspace*{2mm}

\begin{figure}[h!]
\begin{center} 
\includegraphics[scale=0.35, angle=0]{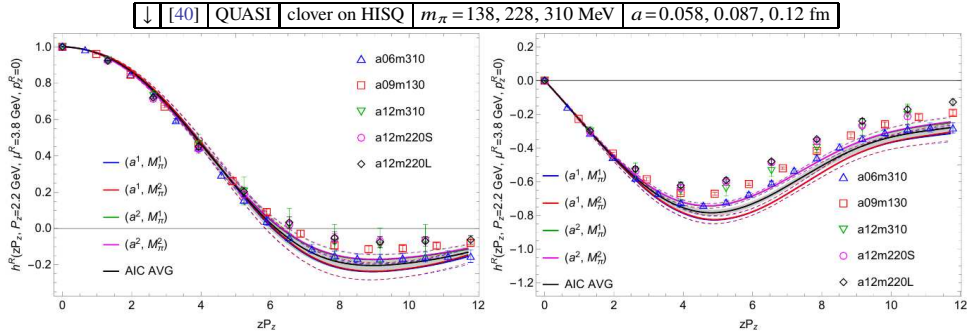}
\setlength{\tabcolsep}{3pt}
\rput(-6.5cm,4.2cm){\scriptsize\begin{tabular}{|c|c|c|c|c|c|}\hline $\downarrow$ & \cite{Lin:2020fsj} & QUASI & clover on HISQ & $m_\pi\!=\!138,228,310$ MeV & $a\!=\!0.058,0.087,0.12$ fm\\\hline\end{tabular}}
\end{center}
\vspace*{-6mm}
\caption{Real (left) and imaginary (right) part of renormalized matrix elements of unpolarized isovector quasi-PDFs \cite{Lin:2020fsj}, for the nucleon boosted to 2.2 GeV. The striking feature is the tiny size of statistical errors, irreconciable with experience of other groups calculating these or similar matrix elements.}
\vspace*{-3mm}
\label{fig:cont3}
\end{figure}

Another study of the continuum limit of unpolarized isovector PDFs was performed in Ref.~\cite{Lin:2020fsj} with a mixed action setup of clover valence on a HISQ sea at three lattice spacings from 0.058 fm to 0.12 fm and three pion masses, including the physical one at the intermediate value of the lattice spacing of 0.087 fm.
Finite volume effects were also tested with three lattice extents at the coarsest lattice spacing.
In principle, the setup allows for a combined continuum limit -- physical pion mass extrapolation.
We illustrate the calculation with renormalized matrix elements at a nucleon boost of 2.2 GeV, see Fig.~\ref{fig:cont3}.
The striking feature of this plot is the tiny magnitude of statistical errors, smaller than the symbol size even at the physical point and at such a high boost.
Results at even higher boosts, up to 3.1 GeV at the physical point, are also included in the analysis.
It is incomprehensible to the author of this review how this could have been achieved.
Such tiny reported errors are in stark contrast with the experience for the signal-to-noise ratio in these matrix elements reported by all other groups.
This experience unambiguosly shows that the signal-to-noise ratio decays exponentially with the nucleon boost, particularly at the physical point.
At the same time, excited-states contamination strongly increases, compromising the extraction of ground state properties, unless one is able to go to very large source-sink separations.
Moreover, even special techniques like distillation (see next subsection) do not seem to be able to obtain such precise results for a highly-boosted nucleon of physical mass.
The typical errors obtained by other groups are summarized in Tab.~1 of Ref.~\cite{Constantinou:2020pek} and are considerably larger even at lower boosts and non-physical pion masses.
Ref.~\cite{Lin:2020fsj} provides no explanation of these tiny errors and thus, no assessment of the final obtained PDFs is possible until serious clarification of the issue.

\subsection{Distillation for PDFs}
As remarked in the previous subsection, the signal for the non-local matrix elements required for extracting partonic distributions decays exponentially with the hadron boost.
The problem is only partially alleviated by the use of momentum smearing \cite{Bali:2016lva}.
This generalization of the standard Gaussian (Wuppertal) smearing \cite{Gusken:1989qx} introduces additional phase factors into the smearing function such that the momentum distribution of quarks becomes centered around some non-zero value.
The method was first explored for quasi-PDFs in Ref.~\cite{Alexandrou:2016jqi}, confirming its high efficiency, and became a standard tool in such computations.
However, it acts no miracles -- larger momenta become accessible, but the exponential signal-to-noise problem still creeps in at some point.
In practice, this problem still hinders robust extraction of partonic functions by setting limits on the achievable boosts, particularly at low pion masses.

An important attempt to circumvent the problem is to combine the momentum smearing technique with distillation.
The latter has been successfully applied in several hadron spectroscopy studies, helping to resolve several states from dense spectra.
It allows for improved volume sampling and a better control of excited states.
In this way, matrix elements can be reliably accessed at smaller source-sink separations, where the signal is exponentially more favorable than at large separations.
The inclusion of momentum smearing into the distillation framework was first explored in Ref.~\cite{Egerer:2020hnc} and applied to PDFs in Ref.~\cite{Egerer:2021ymv}.

\begin{figure}[h!]
\begin{center} 
\includegraphics[scale=0.33, angle=0]{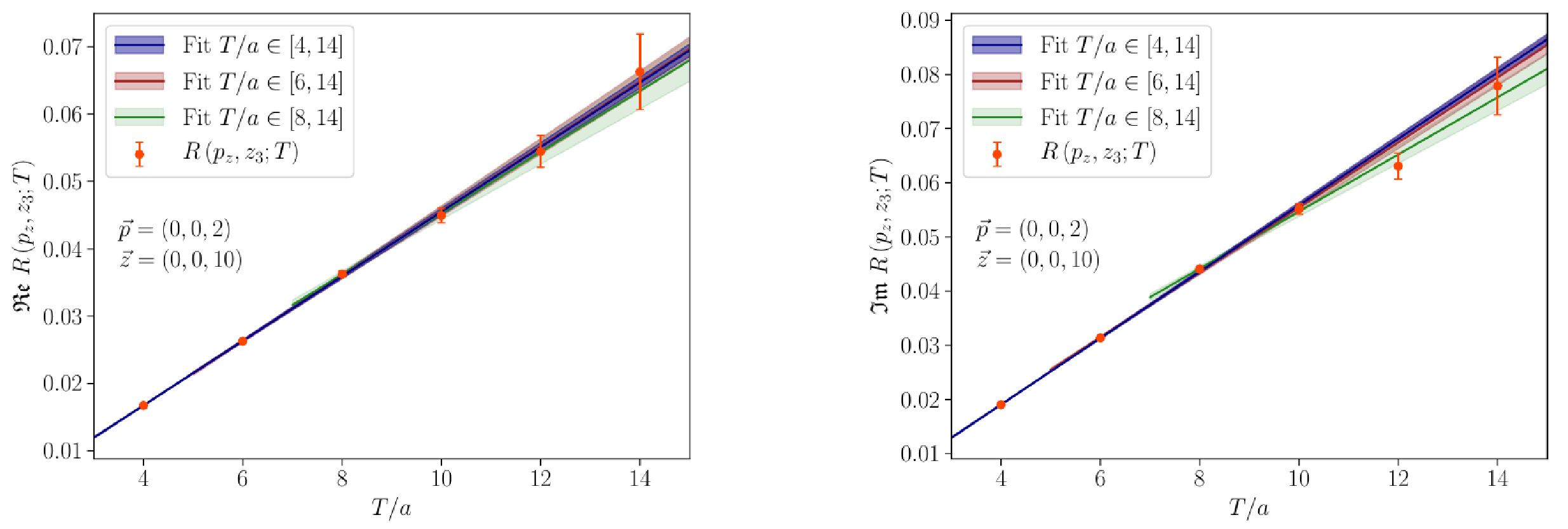}
\includegraphics[scale=0.33, angle=0]{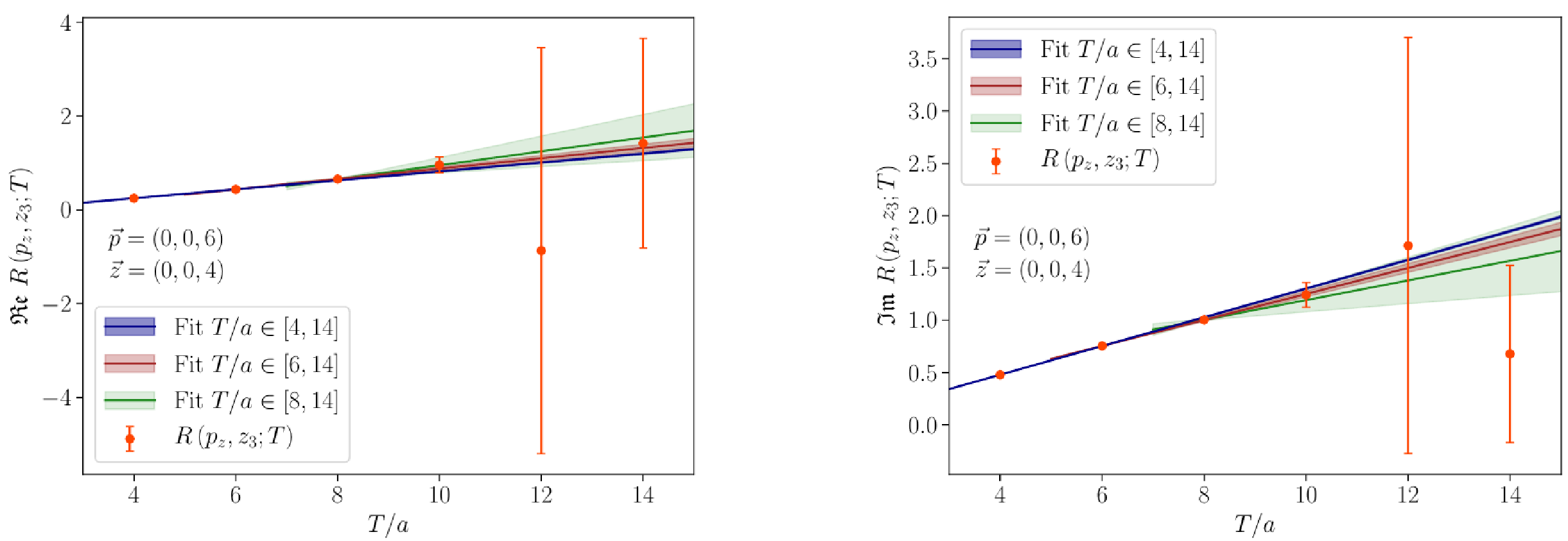}
\setlength{\tabcolsep}{3pt}
\rput(-6.6cm,4.8cm){\scriptsize\begin{tabular}{|c|c|c|c|c|}\hline \cite{Egerer:2021ymv} & PSEUDO & clover & $m_\pi\!=\!358$ MeV & $a\!=\!0.094$ fm\\\hline\end{tabular}}
\end{center}
\vspace*{-5mm}
\caption{The summation method applied to extraction of pseudo-ITDs related to the unpolarized isovector PDFs \cite{Egerer:2021ymv}. The relevant matrix elements are extracted as the slope of the summed ratio of three- and two-point functions for non-local operator insertions from 1 to $T/a-1$, where $T/a$ is the source-sink separation, fitted between varied intervals $[T/a,14]$. The upper (lower) panel represents $z/a=10$ ($z/a=4$) at a nucleon boost of 0.82 (2.47) GeV.}
\vspace*{-3mm}
\label{fig:distillation}
\end{figure}

In the latter work, the HadStruc collaboration used clover fermions at a pion mass of 358 MeV and a lattice spacing of 0.094 fm to extract unpolarized isovector PDFs of the nucleon via the pseudo-distribution approach.
The crucial finding is that distillation with momentum smearing is able to provide pseudo-ITDs of superior quality by profiting from this combination of techniques.
Their robust extraction of the matrix elements is illustrated in Fig.~\ref{fig:distillation}.
The upper panel shows an example of a very successful determination at a relatively low boost.
Note, however, that the used framework allowed for a summation method extraction using as small source-sink separations as 0.038 fm, as compared to required separations of at least 0.8-0.9 fm reported in the ETMC study of Ref.~\cite{Alexandrou:2019lfo}.
The lower panel, in turn, demonstrates a ``difficult'' case of a 2.47 GeV boost, which is apparently on the brink of a robust extraction even with the momentum-smeared distillation approach.
We emphasize again the contrast with the claims of Lin et al.~\cite{Lin:2020fsj}, where much smaller errors are claimed at the physical point with boosts of 2.2-3.1 GeV, using only momentum smearing.\vspace*{1mm}

\begin{figure}[h!]
\begin{center} 
\includegraphics[scale=0.3, angle=0]{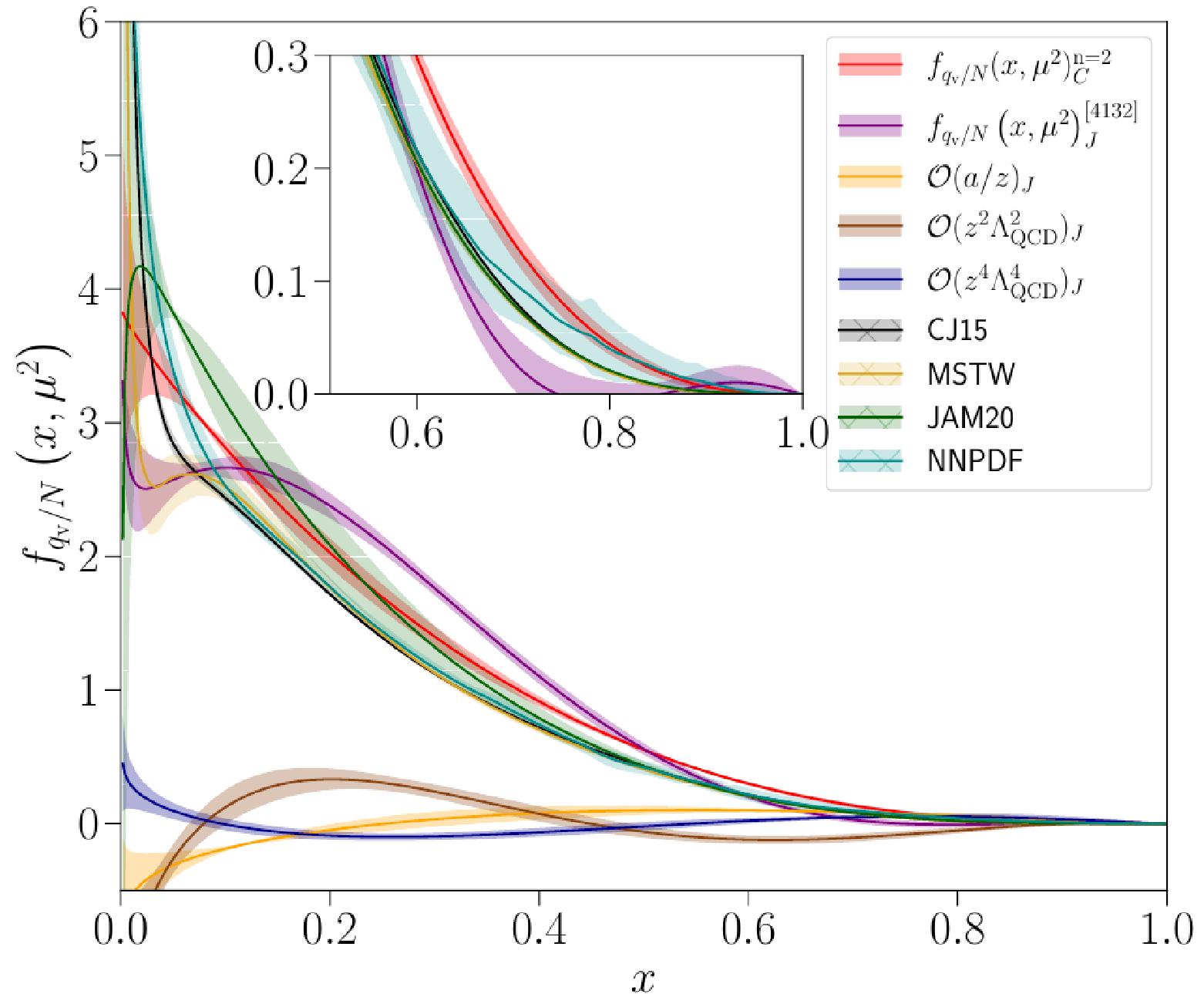}
\includegraphics[scale=0.3, angle=0]{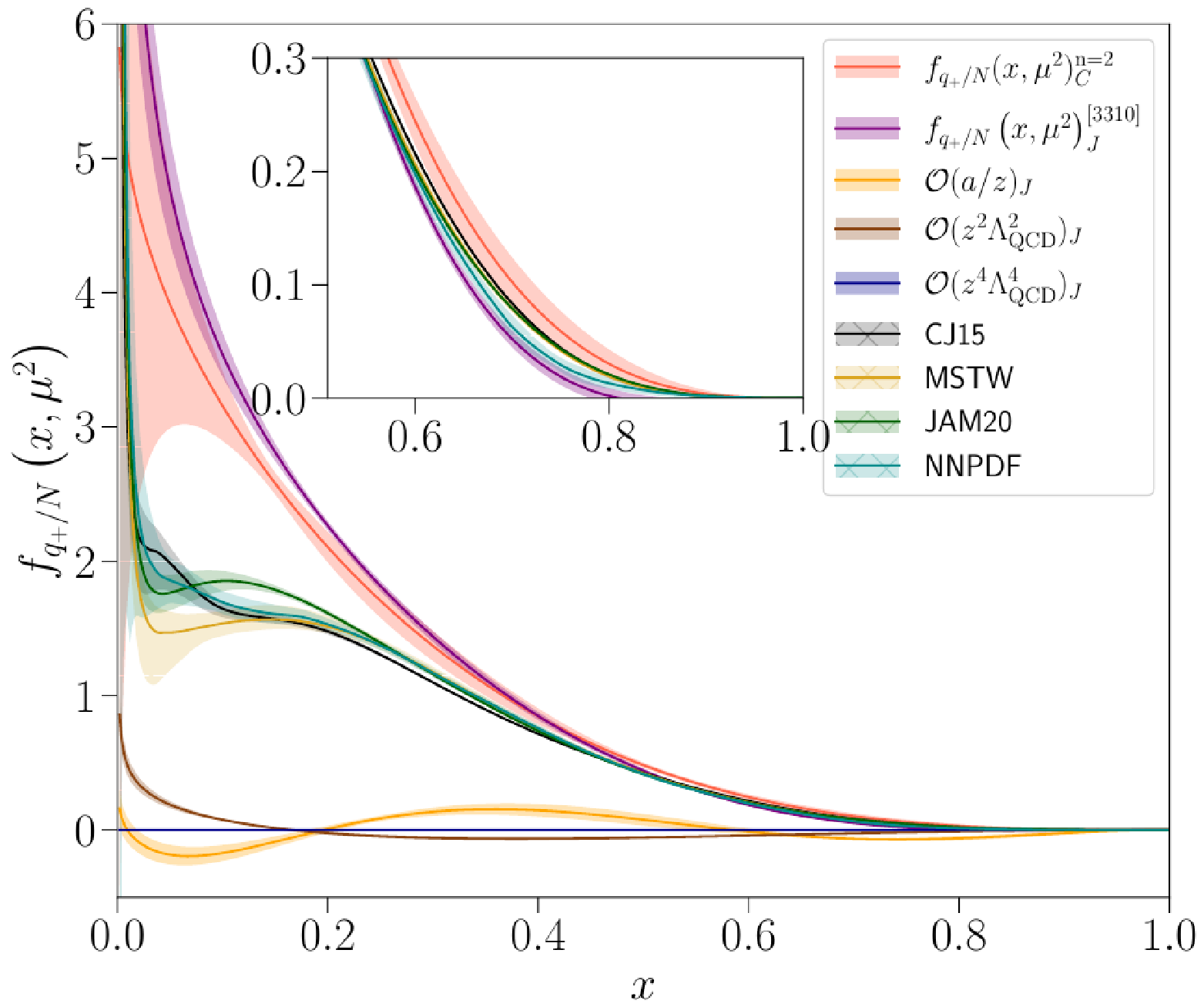}
\setlength{\tabcolsep}{3pt}
\rput(-7.2cm,6.2cm){\scriptsize\begin{tabular}{|c|c|c|c|c|c|}\hline $\downarrow$ & \cite{Egerer:2021ymv} & PSEUDO & clover & $m_\pi\!=\!358$ MeV & $a\!=\!0.094$ fm\\\hline\end{tabular}}
\end{center}
\vspace*{-6mm}
\caption{The final reconstructed unpolarized isovector PDFs of the nucleon from Ref.~\cite{Egerer:2021ymv}. The left panel shows the valence distribution $q_v\equiv q-\bar{q}$ related to the real part of ITDs. The right panel is the distribution $q_+\equiv q+\bar{q}$ related to the imaginary part. The results of the lattice analysis are shown with red (purple) bands for the uncorrelated (correlated) fits. Also pictured are discretization effects (yellow), twist-4 (brown) and twist-6 HTE (blue) and a selection of global fits.}
\vspace*{-3mm}
\label{fig:distillation2}
\end{figure}

Apart from demonstrating the efficiency of the distillation technique for PDFs, an important development in Ref.~\cite{Egerer:2021ymv} is also a consideration of the role of correlations between the lattice data and of discretization and higher-twist effects.
This part utilized the framework of Jacobi polynomials for systematic uncertainties introduced in Ref.~\cite{Karpie:2021pap}.
We refer the Reader to the original reference for the details of this rather complicated analysis and summarize here the conclusions.
Most importantly, neglecting correlations can lead to erroneous final distributions, represented by the red bands in Fig.~\ref{fig:distillation2}.
The correlations are properly included in the purple bands, obtained from a variant of Jacoby-polynomial Bayesian fits, including discretization effects and twist-4,6 HTE.
It was also found that only the inclusion of discretization effects leads to consistency of the lattice data with the DGLAP evolution.
Note that the correlations-including PDFs are further away from the global fits than the uncorrelated ones, particularly for the valence distribution.
Thus, not taking correlations into account may lead to an overoptimistic, but erroneous conclusion of a rather good agreement with phenomenology.
Obviously, such an agreement should not be expected at a non-physical pion mass of this study and with yet unaccounted for systematics.

The success of the distillation framework for partonic distributions inspired also additional work of the HadStruc collaboration, concerning nucleon's gluon PDFs (see next subsection) \cite{HadStruc:2021wmh,Sufian:LAT21} and transversity PDFs \cite{Karthik:LAT21}.

\subsection{Gluon PDFs}
\begin{figure}[h!]
\begin{center} 
\includegraphics[scale=0.24, angle=0]{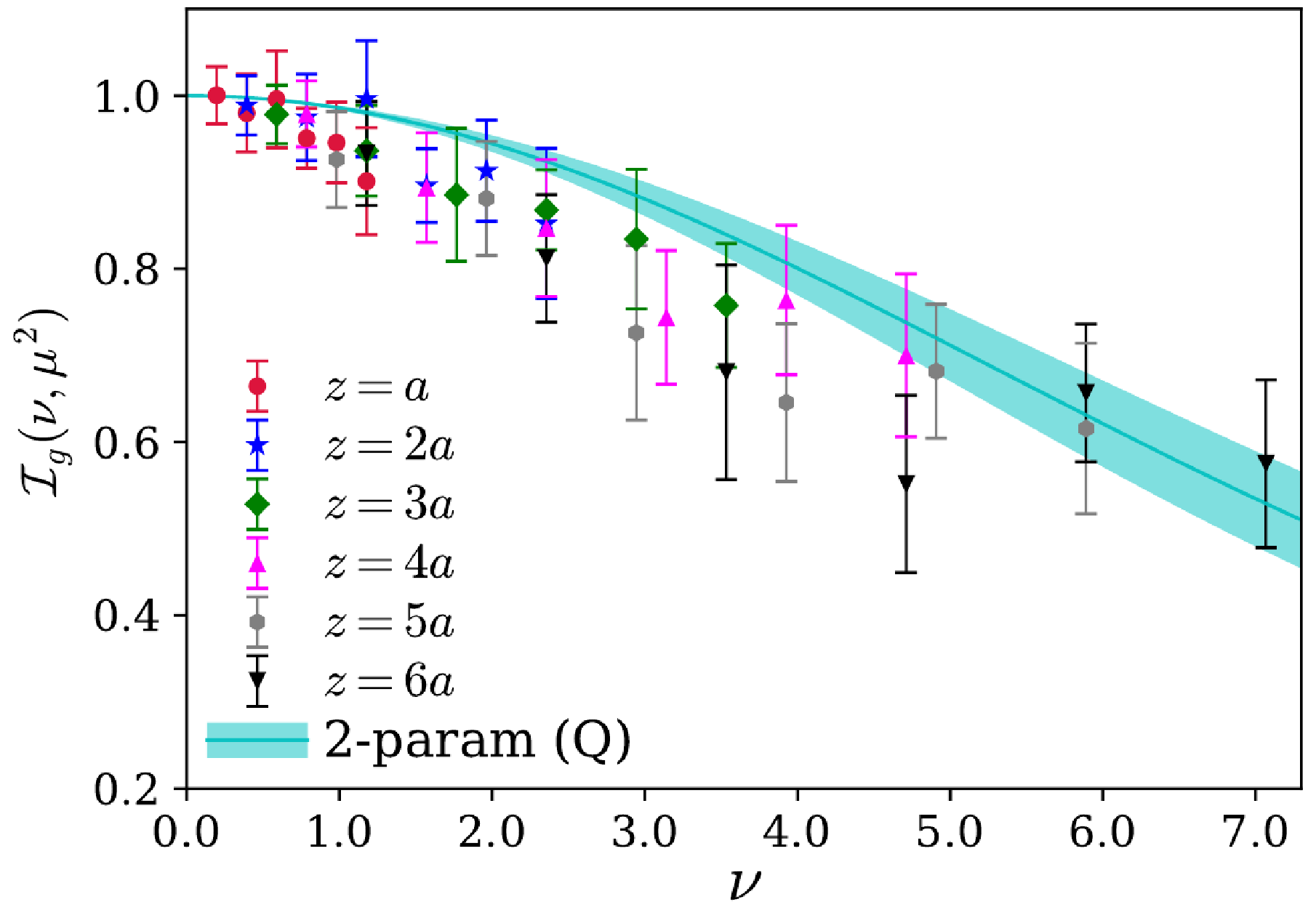}
\includegraphics[scale=0.258, angle=0]{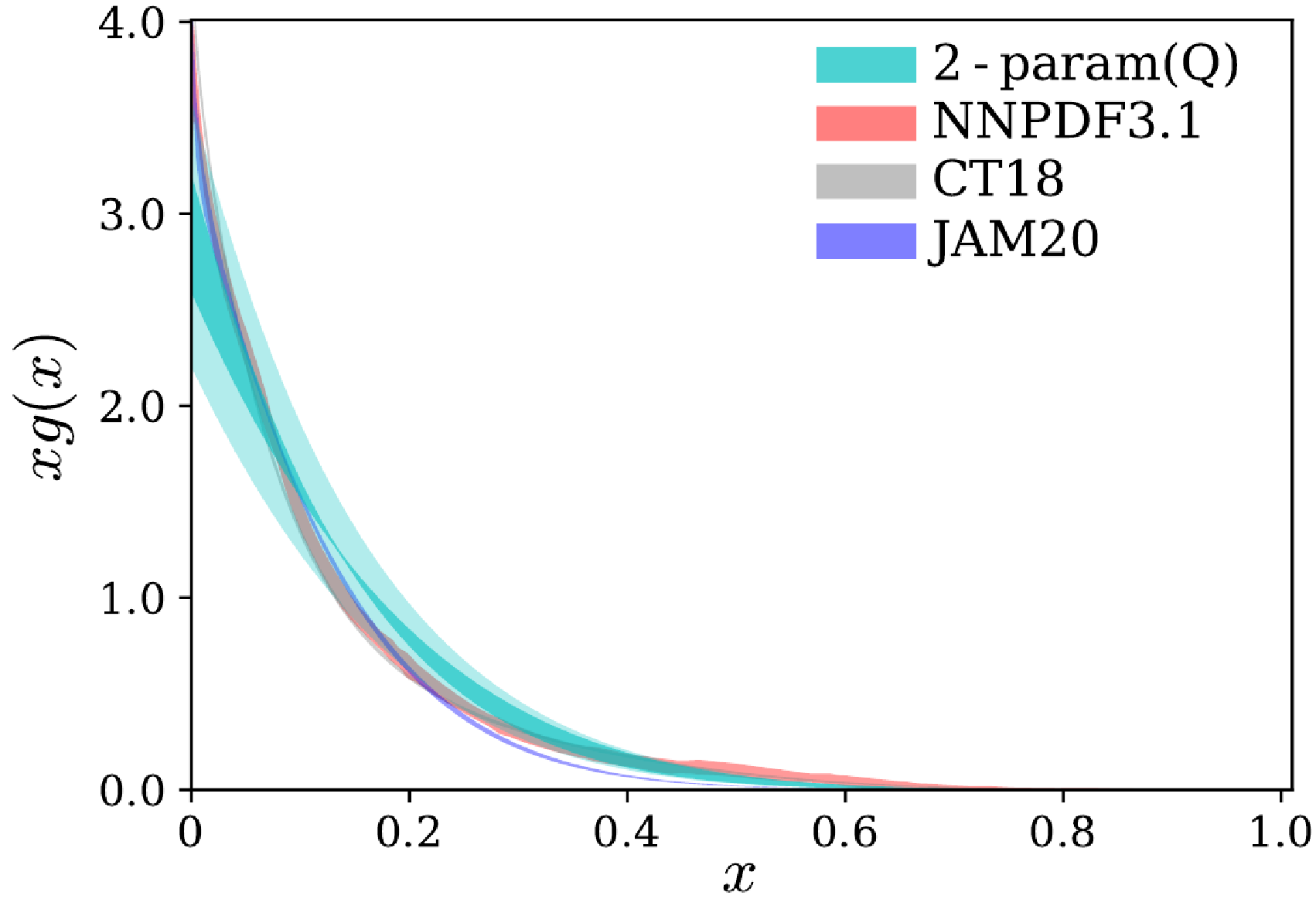}
\setlength{\tabcolsep}{3pt}
\rput(-7cm,5.35cm){\scriptsize\begin{tabular}{|c|c|c|c|c|c|}\hline $\downarrow$ & \cite{HadStruc:2021wmh} & PSEUDO & clover & $m_\pi\!=\!358$ MeV & $a\!=\!0.094$ fm\\\hline\end{tabular}}
\end{center}
\vspace*{-5mm}
\caption{Gluon PDF via the pseudo-distribution approach from the HadStruc collaboration \cite{HadStruc:2021wmh}. The left panel shows matched ITDs together with a two-parameter fit to reconstruct the PDF. The right panel compares the reconstructed PDF (cyan band with its darker part representing the statistical error and the lighter one the total uncertainty after normalizing with the gluon momentum fraction from an independent calculation) with selected global fits.}
\label{fig:gluon}
\end{figure}

Typically, gluonic quantities in lattice hadron structure suffer from rather unfavorable signal and thus, gluon PDFs are considerably difficult to address.
Recently, a robust exploratory study emerged by the HadStruc collaboration \cite{HadStruc:2021wmh}, wherein they combined the momentum-smeared distillation technique with the summed GEVP method of extracting the matrix elements, profiting additionally from the extended basis of interpolators afforded by distillation. 
Gradient flow was applied to suppress UV fluctuations, with physics extracted via a zero flow time limit.
The combination of these techniques allowed for a good control over the signal.
The final PDFs, see Fig.~\ref{fig:gluon}, were reconstructed using Jacobi polynomials, after matching of pseudo- to light-cone ITDs upon neglecting the mixing with singlet quark PDFs and fixing the global normalization via the gluon momentum fraction from an independent calculation.
The outcome was compared to phenomenological determinations and very reasonable agreement was found in the whole $x$-range.
Clearly and typically for all lattice calculations of partonic functions, several sources of systematics need to be investigated.
However, the already excellent agreement with global fits justifies the conclusion that very likely these systematic effects are not overwhelming, in particular the neglected mixing may be a small effect.
We also note the very good statistical quality of the results, with the bulk of the total error at small $x$ coming from the normalization.
Preliminary results for the polarized gluon PDF were also presented in the conference \cite{Sufian:LAT21}.

\begin{figure}[h!]
\begin{center}
\hspace*{6.7cm}
\includegraphics[scale=0.2, angle=0]{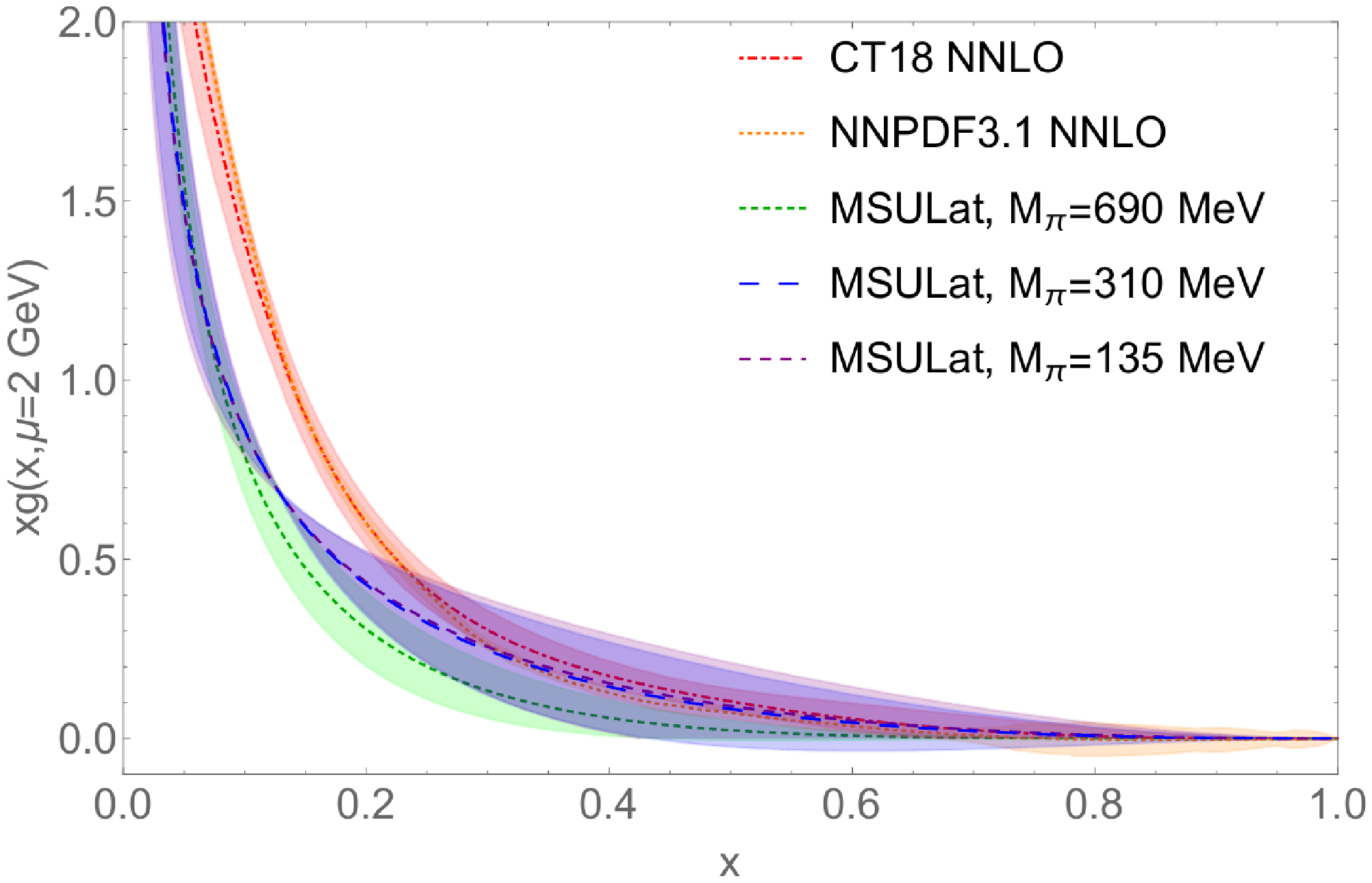}
\rput(-10.5cm,3.21cm){\includegraphics[scale=0.25, angle=0]{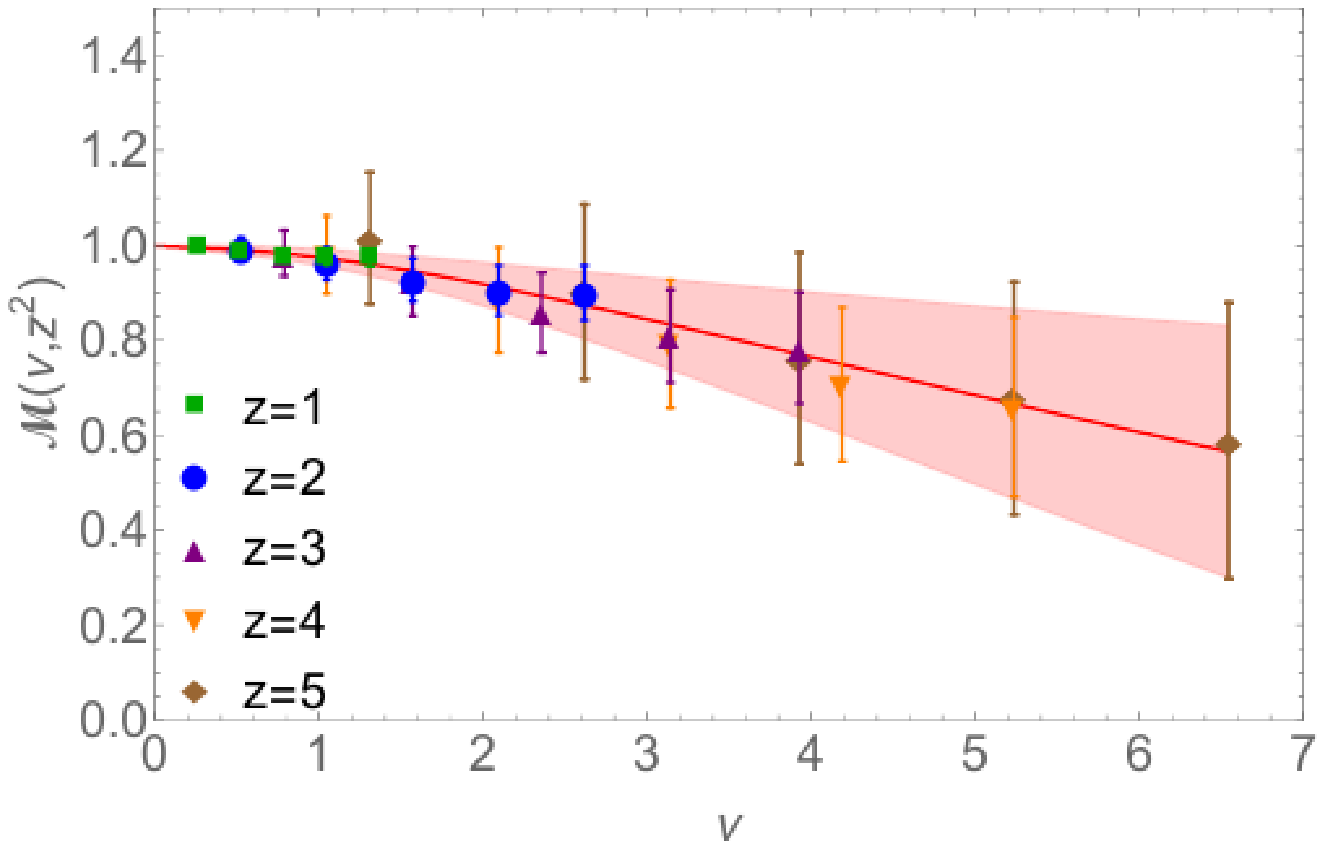}}
\rput(-10.5cm,1.03cm){\includegraphics[scale=0.25, angle=0]{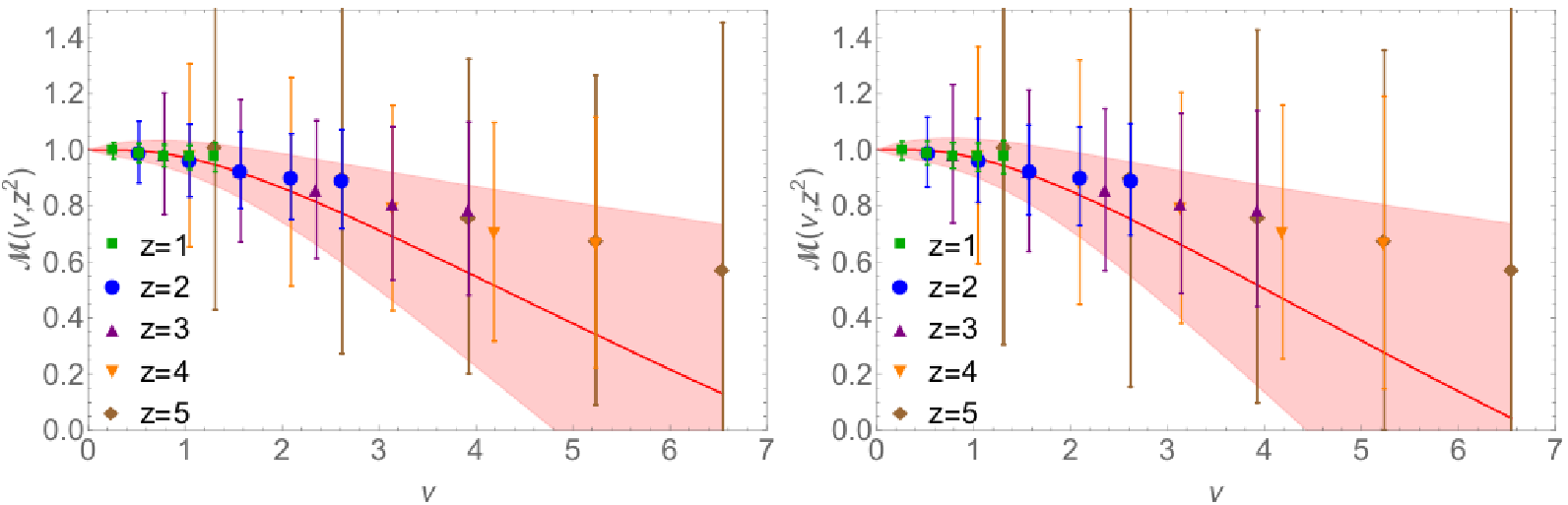}}
\setlength{\tabcolsep}{3pt}
\rput(-6.7cm,4.6cm){\scriptsize\begin{tabular}{|c|c|c|c|c|c|}\hline $\downarrow$ & \cite{Fan:2020cpa} & PSEUDO & clover on HISQ & $m_\pi^{\rm sea}=310$ MeV, $m_\pi^{\rm val}\!=\!310,690$ MeV & $a\!=\!0.12$ fm\\\hline\end{tabular}}
\end{center}
\vspace*{-5mm}
\caption{Gluon PDF via the pseudo-distribution approach from the East Lansing group \cite{Fan:2020cpa}. The left part shows reduced ITDs together with an $z$-expansion fit to reconstruct the PDF (top: $m_\pi^{\rm val}\!=\!690$ MeV, bottom left: $m_\pi^{\rm val}\!=\!310$ MeV, bottom right: extrapolated to the physical point). The right part compares the reconstructed PDF (green band: $m_\pi^{\rm val}\!=\!690$ MeV, blue band: $m_\pi^{\rm val}\!=\!310$ MeV, purple band: extrapolated to the physical point) with selected global fits.}
\label{fig:gluon2}
\vspace*{-3mm}
\end{figure}

Another calculation of gluon PDFs via the pseudo-distribution approach was performed recently by the East Lansing group \cite{Fan:2020cpa}, with a mixed action setup of clover valence on HISQ sea quarks.
The valence quark mass was set to two values and an extrapolation to the physical pion mass was attempted.
In Fig.~\ref{fig:gluon2}, shown are the reduced ITDs for the two masses together with the extrapolation and the final gluon PDF.
Rather strikingly, the result quoted as the physical point is basically indistinguishable from the one of the lighter valence pion mass, both for its central value and its error.
The final gluon PDF is in much worse agreement with the global fits for small $x$, as compared to the HadStruc computation, being significantly below CT18 and NNPDF3.1.
Obviously, systematic uncertainties need to be carefully scrutinized.
However, the influence of one of these, the mixing with the isoscalar quark PDF, has already been estimated at 4\% by taking the CT18 singlet quark PDF values.
Thus, the current discrepancy with global fits must be driven by other sources.
The pseudo-PDF approach and the same lattice setup with an additional lighter valence mass and an additional coarser lattice spacing at the intermediate mass was also used to explore gluon PDFs in the pion \cite{Fan:2021bcr}.

\subsection{Flavor-singlet PDFs}
The quark PDFs computations reported above were all for the flavor non-singlet $u-d$ combination. 
However, ultimately one is interested also in the decomposition of the $u-d$ combination into the separate $u$ and $d$ parts, as well as in obtaining the PDFs of the heavier quarks.
Obviously, flavor-singlet quark PDFs are much more demanding, since they require the calculation of the noisy quark-disconnected diagrams and moreover, they are subject to mixing with the gluon PDFs.
Regarding the latter, the situation is somewhat different for the light and heavier sea quarks in the nucleon.
Since up and down quarks are valence partons, the mixing with the gluon can be reasonably conjectured to be a small effect, likely subleading at the current precision level of lattice partonic functions.
In turn, strange and charm quarks appear in the nucleon predominantly from gluon splitting and hence, mixing with gluon PDFs may be a sizable effect.

\begin{figure}[h!]
\begin{center} 
\includegraphics[scale=0.36, angle=0]{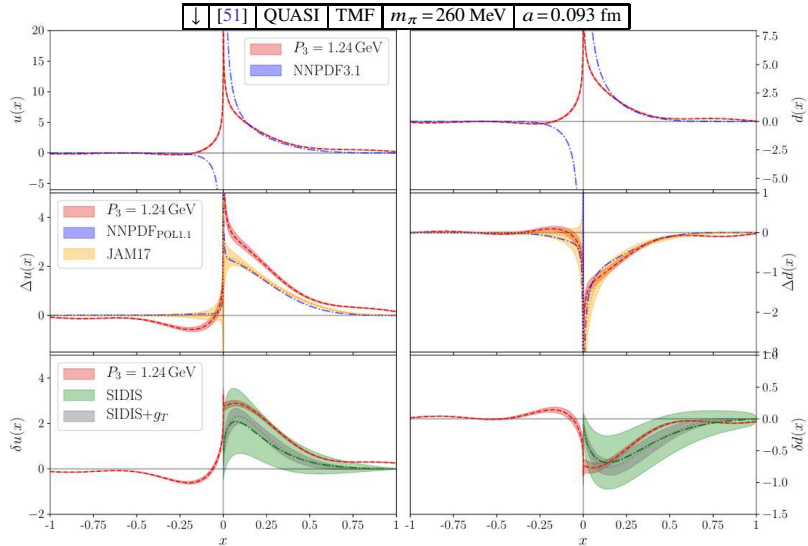}
\setlength{\tabcolsep}{3pt}
\rput(-5.5cm,7.05cm){\scriptsize\begin{tabular}{|c|c|c|c|c|c|}\hline $\downarrow$ & \cite{Alexandrou:2021oih} & QUASI & TMF & $m_\pi\!=\!260$ MeV & $a\!=\!0.093$ fm\\\hline\end{tabular}}
\end{center}
\vspace*{-5mm}
\caption{ETMC's flavor-decomposed nucleon quark PDFs (left: $u$, right: $d$) from the quasi-distribution approach \cite{Alexandrou:2021oih}. From top to bottom: unpolarized, helicity, transversity compared to selected global fits.}
\vspace*{-3mm}
\label{fig:decomp}
\end{figure}

The first extraction of flavor-decomposed PDFs was done by ETMC for the helicity case \cite{Alexandrou:2020uyt}. The most computationally challenging part of evaluating the disconnected contributions was done utilizing techniques proven before to be successful for local operators, including explicit treatment of the low modes and stochastic evaluation of the high modes using hierarchical probing and the one-end trick.
The work was soon extended to the other types of PDFs \cite{Alexandrou:2021oih}, with all light-quark distributions shown in Fig.~\ref{fig:decomp} (neglected mixing with the gluon sector).
We note considerably good agreement with selected phenomenological fits, with quantitatively robust conclusions to be established upon investigation of systematics.
Results for the strange quark were also calculated, with the above reservation that the ignored mixing with the gluon is a more crude approximation than for the light quarks.

While the flavor-decomposed PDFs for the strange quark may be considered a preliminary addition to Refs.~\cite{Alexandrou:2020uyt,Alexandrou:2021oih}, results for the strange and charm quark PDFs are the only ones presented by the East Lansing group in Ref.~\cite{Zhang:2020dkn}.
The authors used the mixed action setup identical to the one of their computation of gluon pseudo-PDFs of the nucleon and attempted to draw physical conclusions about the strange-antistrange and strange-anticharm symmetry.
In the light of the above comment about the ignored gluon splitting, such conclusions can be considered premature.

\subsection{Generalized parton distributions (GPDs)}
GPDs are one of the key quantities in probing the three-dimensional structure of the nucleon. Yet, they are presently rather poorly constrained by experiment and hence, first-principle insights from LQCD may play a crucial role.
The first lattice extraction of nucleon GPDs appeared in 2020 from the quasi-distribution approach \cite{Alexandrou:2020uyt}.
ETMC used a heavier-than-physical pion mass ensemble in their exploratory study of unpolarized and helicity GPDs, both at zero (with only transverse momentum transfer, $Q$) and non-zero skewness (also with longitudinal $Q$).
It is worthwhile to realize the additional difficulties encountered in GPDs calculations, as compared to PDFs.
One of the lattice challenges is further decreasing signal quality with increasing $Q^2$.
Additionally, working with the standard GPDs defined in the Breit frame, one needs separate inversions for each $Q$, with carefully tuned momentum smearing parameters to ensure optimal signal.
Moreover, in the chiral-even (odd) case, there are two (four) twist-2 GPDs,\footnote{The standard notation is for the chiral-even cases: $H,\,E$ (unpolarized), $\tilde{H},\,\tilde{E}$ (helicity); for the chiral-odd case: $H_T,\,E_T,\,\tilde{H}_T,\,\tilde{E}_T$ (transversity).} which necessitates projecting the 3-point functions with two (four) projectors to disentangle the different GPDs.
Effectively, this also lowers the signal quality, as some projectors are more noisy.\vspace*{1mm}

\begin{figure}[h!]
\begin{center} 
\includegraphics[scale=0.51, angle=0]{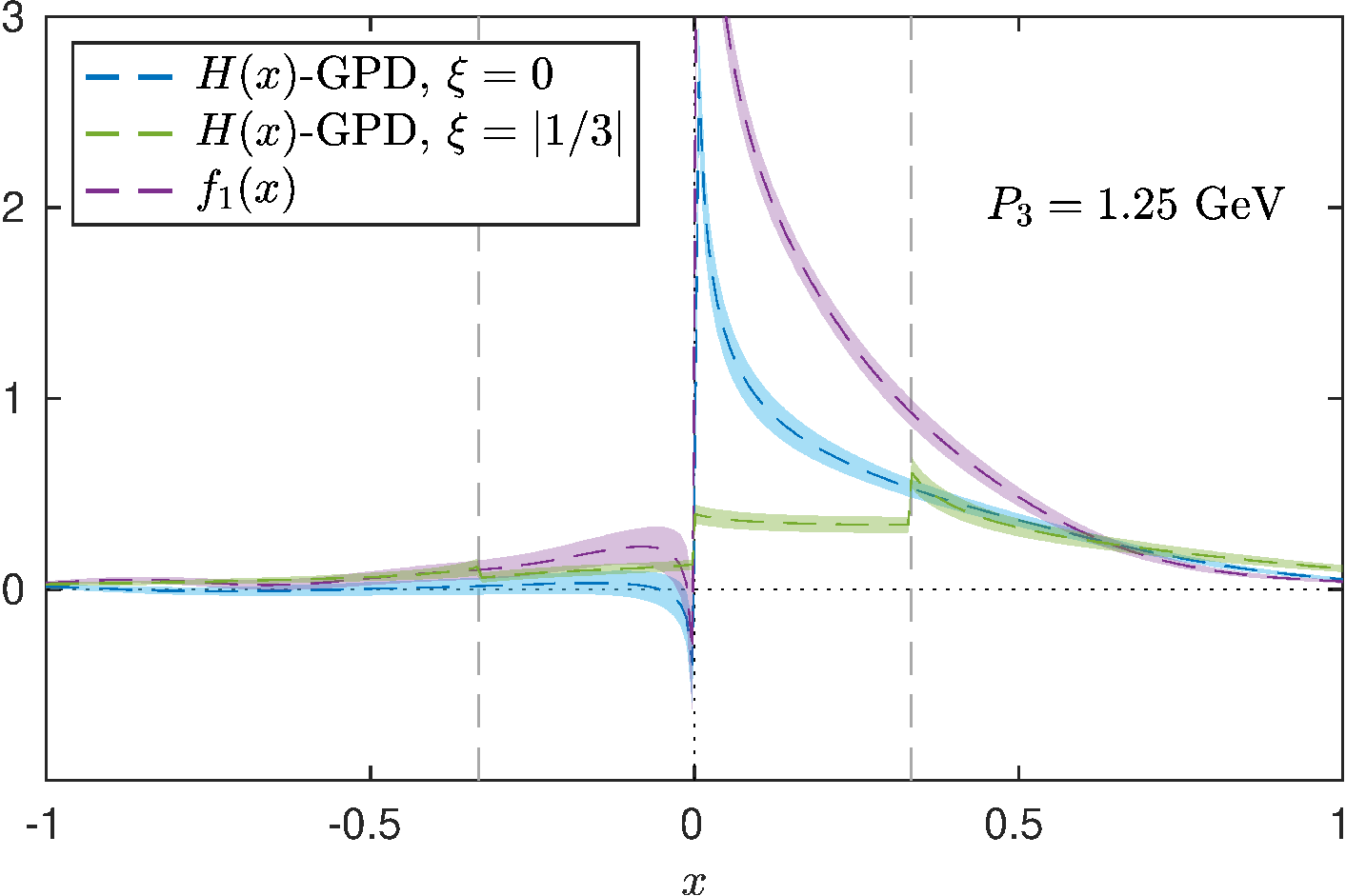}
\includegraphics[scale=0.51, angle=0]{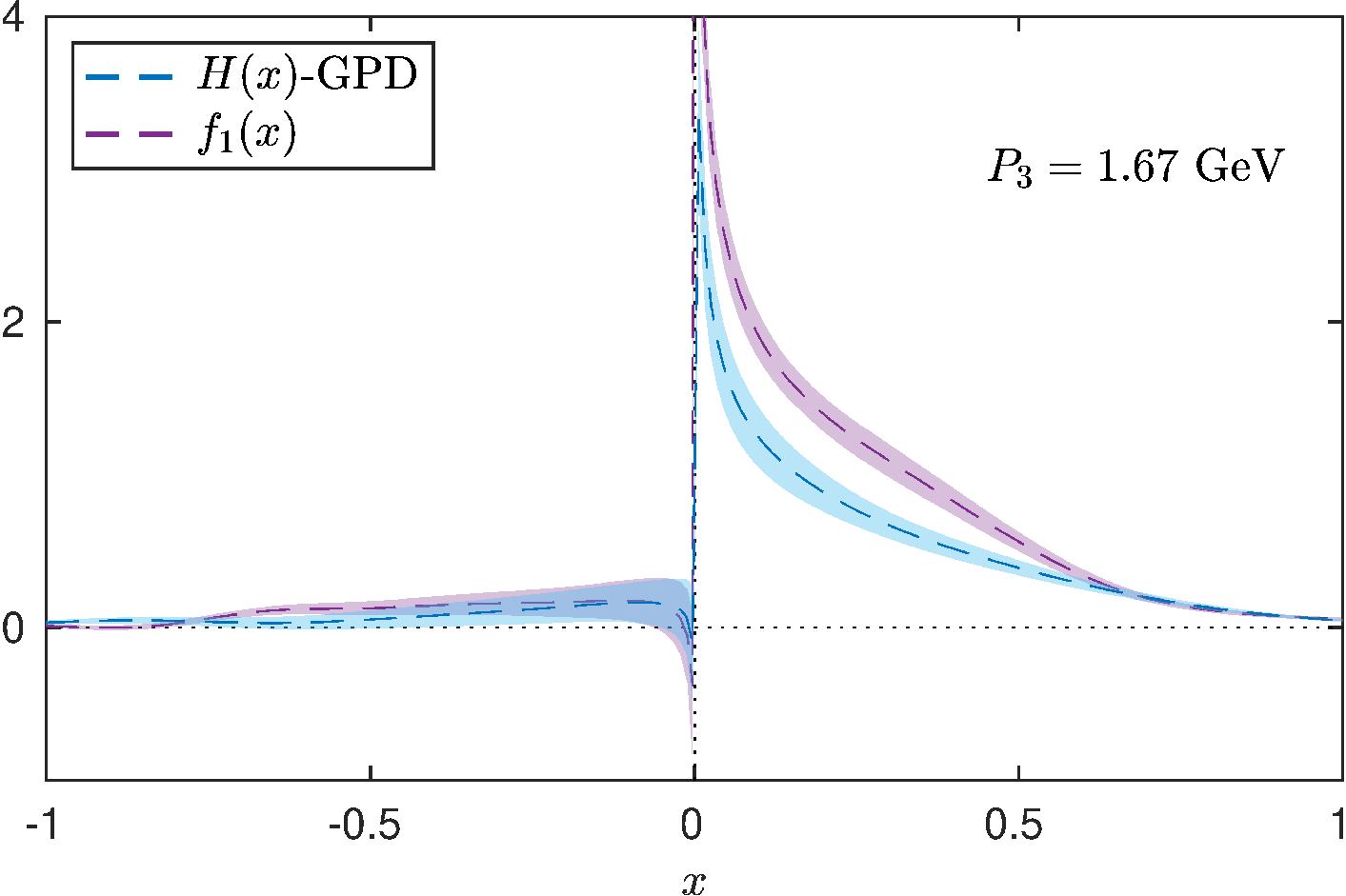}
\setlength{\tabcolsep}{3pt}
\rput(-7cm,5.08cm){\scriptsize\begin{tabular}{|c|c|c|c|c|c|}\hline $\downarrow$ & \cite{Alexandrou:2020uyt} & QUASI & TMF & $m_\pi\!=\!260$ MeV & $a\!=\!0.093$ fm\\\hline\end{tabular}}
\end{center}
\vspace*{-6mm}
\caption{Comparison of unpolarized PDF ($f_1$) with its off-forward generalization ($H$-GPD), from an ETMC calculation \cite{Alexandrou:2020uyt} via the quasi-distribution approach. Left: nucleon boost $P_3=1.25$ GeV, right: $P_3$=1.67 GeV. The $\xi=0$ ($\xi=1/3$) data corresponds to $Q^2=0.69$ ($1.02$) GeV$^2$.}
\vspace*{-3mm}
\label{fig:GPDs}
\end{figure}

Example results for unpolarized GPDs are shown in Fig.~\ref{fig:GPDs} at two nucleon boosts, as compared to the corresponding PDFs.
As expected, the GPDs are suppressed with respect to PDFs, particularly at small $x$.
In the non-zero skewness case ($\xi=1/3$), attainable only at the lower boost, two distinct regions appear, the standard Dokshitzer-
Gribov-Lipatov-Altarelli-Parisi (DGLAP; $|x|>\xi$) region and the Efremov-Radyushkin-Brodsky-Lepage (ERBL; $|x|<\xi$) region.
In the latter, the GPDs are further suppressed.
Note, however, that the observed discontinuity at the border between the regions is a manifestation of enhanced HTE, expected from model calculations \cite{Bhattacharya:2018zxi,Bhattacharya:2019cme}.
The ETMC work was very recently extended to the chiral-odd transversity case with four GPDs \cite{Alexandrou:2021bbo}, using the same lattice setup.

Independent work for GPDs was reported by Lin \cite{Lin:2020rxa} shortly after the ETMC study, using a physical pion mass HISQ ensemble ($a=0.09$ fm) with clover valence fermions.
In the initial version of the paper, non-standard definition (non-Breit frame) of GPDs was employed, with a change to Breit frame done in v2 on arXiv.
Similarly to Ref.~\cite{Lin:2020fsj}, the author achieved surprisingly clean statistical signal for the pertinent matrix elements, with errors as small as $\approx\!2\%$.
Such tiny errors are obtained despite boosting the nucleon to 2.2 GeV, working with physically light quarks and in the presence of momentum transfer further lowering the signal quality.
Comparing the signal more quantitatively, the current statistical errors and number of measurements of Ref.~\cite{Alexandrou:2020uyt} indicate that such errors would be achieved in the ETMC setup with statistics larger by a factor of $\mathcal{O}(20)$ than the one reported by Lin and that is with a 260 MeV pion mass and a boost of only 1.67 GeV.
Extrapolating the cost to the case with one more unit of nucleon boost ($\approx\!\!2.1$ GeV) extends the contrast to a factor of a few hundred in statistics, still at a twice heavier-than-physical pion mass.
Furthermore, it is instructive to compare the signal reported in Ref.~\cite{Lin:2020rxa} with the one obtained for quasi-GPDs of the pion \cite{Chen:2019lcm} by the same author with additional collaborators and in a similar setup of clover valence quarks on a HISQ sea (coarser lattice spacing, $a=0.12$ fm).
The pion work was done with boosts extending only to around 1.7 GeV and at a non-physical pion mass of 310 MeV.
As expected, the quality of the signal is roughly comparable to ETMC's work on nucleon GPDs.
However, astonishingly, the signal claimed in Ref.~\cite{Lin:2020rxa}, at considerably larger boosts and at a more than twice smaller pion mass, is an order of magnitude better, with only about twice larger statistics than in Ref.~\cite{Chen:2019lcm}.
In the opinion of the author of this review, such striking contrasts demand an explanation, even more so with similar signal quality discrepancies observed with respect to several other studies of partonic distributions.
In the present situation, it is, thus, difficult to assess the robustness of results reported by Lin.

\subsection{Twist-3 PDFs/GPDs}
Partonic distributions can be classified according to their twist, which describes the order in the inverse energy scale of the process at which they appear in factorization formulae.
Leading twist (twist-2) functions are the most important for kinematical reasons, but the interest in higher-twist distributions is increasing, given their importance for obtaining the full picture of hadron structure.
While twist-3 PDFs have no probability density interpretation, they appear in QCD factorization theorems for a variety of hard scattering processes and contain crucial information about quark-gluon-quark correlations.
Their experimental measurements are difficult (but planned, e.g.\ at the EIC) and hence, a significant role in their determination can be foreseen for the lattice.
There are three twist-3 PDFs, the chiral-even $g_T(x)$ and the chiral-odd $h_L(x)$ and $e(x)$.\footnote{Their twist-2 counterparts are, respectively, the helicity PDF $g_1(x)$, the transversity PDF $h_1(x)$ and no counterpart for the scalar $e(x)$.}\vspace*{1mm}

\begin{figure}[h!]
\begin{center} 
\includegraphics[scale=0.51, angle=0]{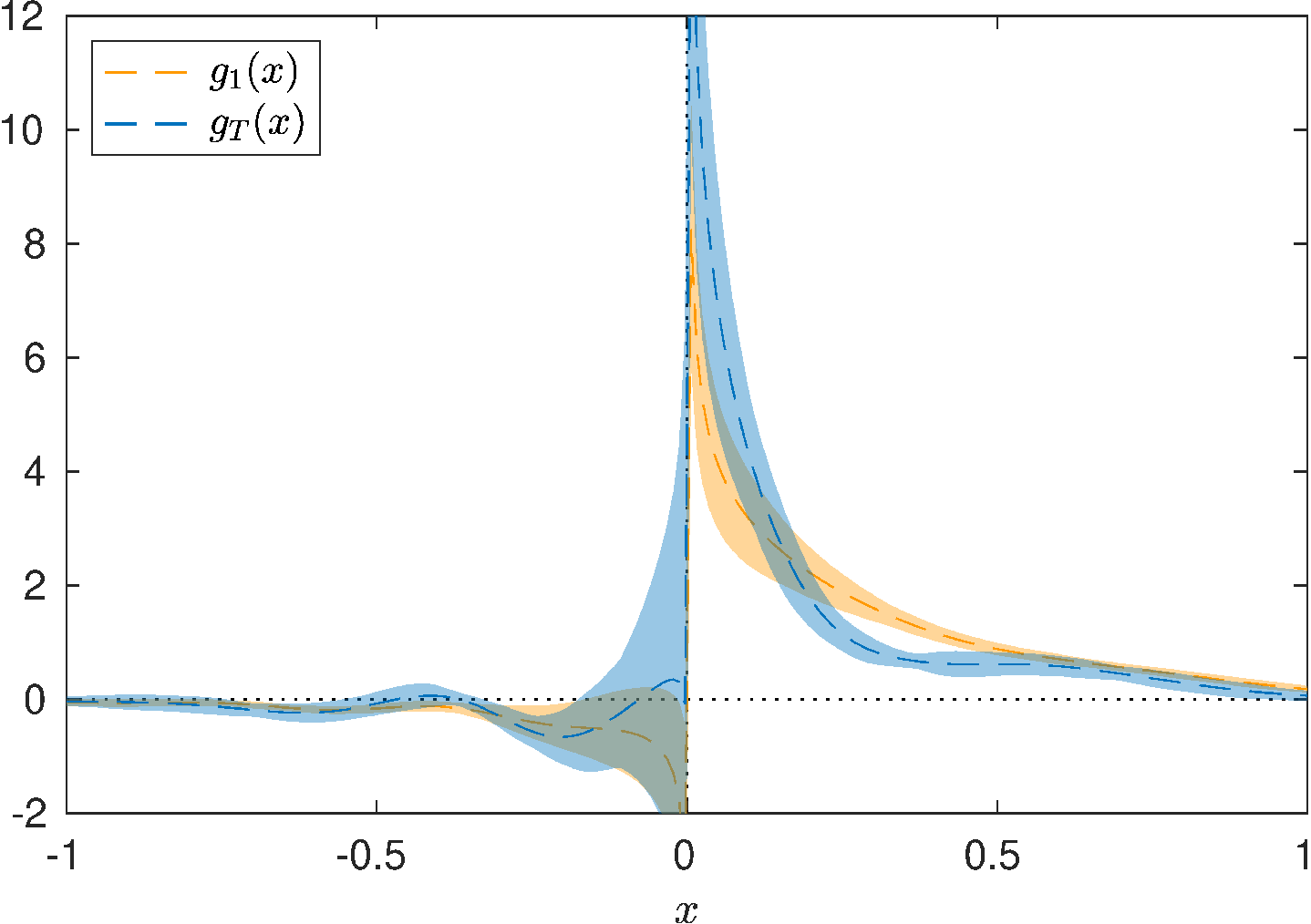}
\includegraphics[scale=0.51, angle=0]{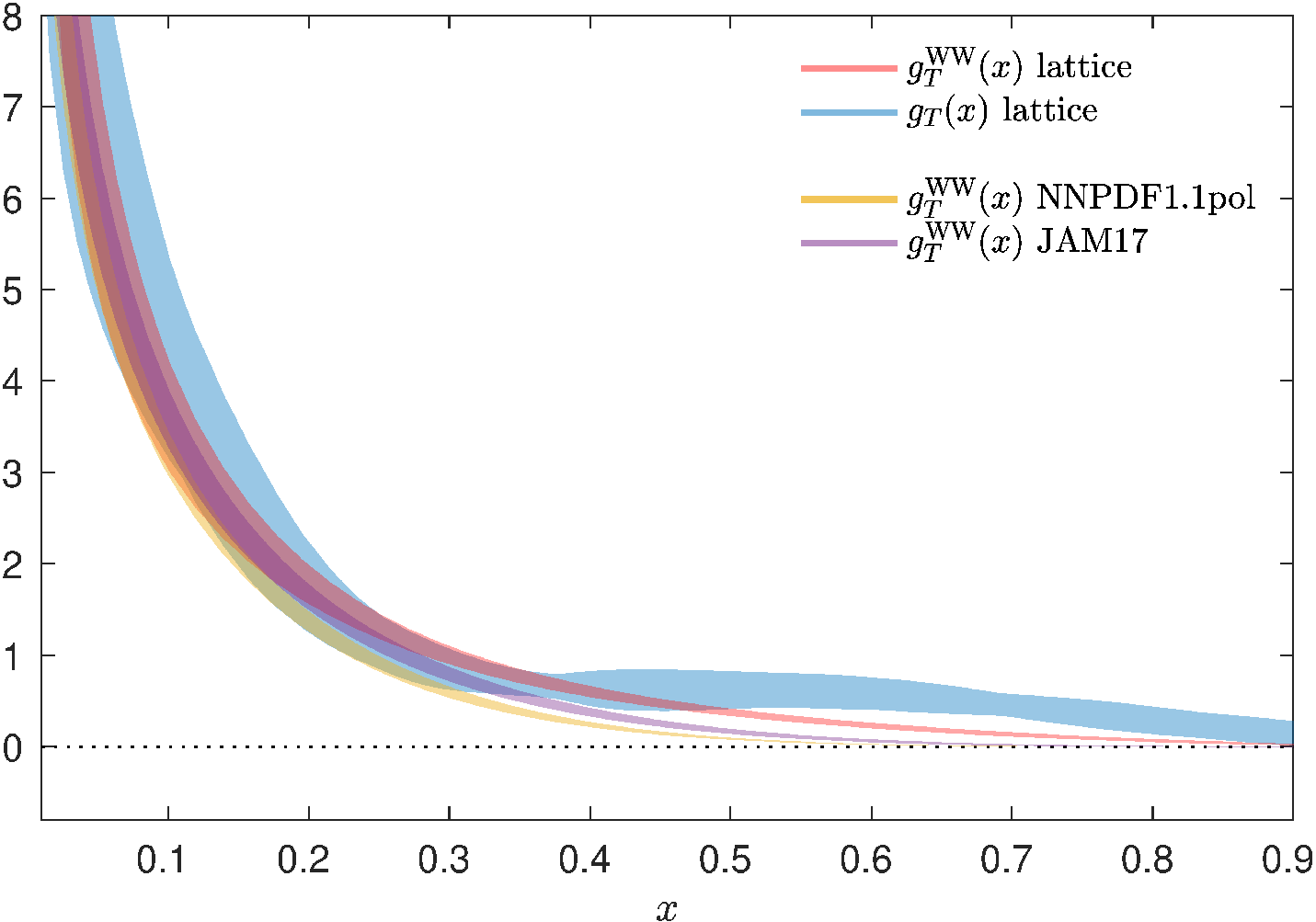}
\setlength{\tabcolsep}{3pt}
\rput(-7.2cm,5.4cm){\scriptsize\begin{tabular}{|c|c|c|c|c|c|}\hline $\downarrow$ & \cite{Bhattacharya:2020cen} & QUASI & TMF & $m_\pi\!=\!260$ MeV & $a\!=\!0.093$ fm\\\hline\end{tabular}}
\end{center}
\vspace*{-5mm}
\caption{Left: comparison of the twist-3 PDF $g_T$ with its twist-2 counterpart $g_1$ \cite{Bhattacharya:2020cen}.
Right: test of the Wandzura-Wilczek approximation for $g_T$.}
\vspace*{-3mm}
\label{fig:twist3}
\end{figure}

First exploratory studies of twist-3 functions were initiated last year in a series of papers by Bhattacharya et al.~\cite{Bhattacharya:2020xlt,Bhattacharya:2020cen,Bhattacharya:2020jfj,Bhattacharya:2021moj}.
In Ref.~\cite{Bhattacharya:2020xlt}, 1-loop matching was derived for the first time for a twist-3 case, the $g_T(x)$ function.
A lattice computation utilizing this matching appeared in Ref.~\cite{Bhattacharya:2020cen}.
The left panel of Fig.~\ref{fig:twist3} shows a comparison of the extracted $g_T$ with $g_1$.
Both functions are similar in magnitude, with a somewhat steeper descent of the twist-3 one at small $x$.
In the right panel, the celebrated Wandzura-Wilczek (WW) approximation \cite{Wandzura:1977qf} was tested, consisting in the twist-3 $g_T$ function being fully determined by $g_1$, i.e.\ $g_T(x)\approx g_T^{\rm WW}(x)\equiv\int_x^1 g_1(y)\,dy/y$.
The WW relation appears to hold for $x\lesssim0.5$, although within rather large uncertainties (and unquantified systematics) and, thus, within possible violations similar to ones found within global analyses \cite{Accardi:2009au}, i.e.\ up to around 40\%.
The matching for the chiral-odd twist-3 functions was addressed in Ref.~\cite{Bhattacharya:2020jfj}, with a special attention paid to the role of zero-mode contributions.
The $h_L$ function was very recently explored numerically \cite{Bhattacharya:2021moj}, with qualitatively very similar conclusions to the ones for $g_T$, including the validity of the WW approximation.
Even more recently, first results appeared \cite{Bhattacharya:2021rua} for the twist-3 GPD $\tilde{G}_2(x)$, whose combination with the helicity twist-2 GPD $\tilde{H}(x)$ is the off-forward generalization of $g_T(x)$.

It is important to emphasize that the above exploratory studies have not taken into account mixing with quark-gluon-quark ($qgq$) operators.
Their role was highlighted in Refs.~\cite{Braun:2021aon} ($g_T$) and \cite{Braun:2021gsk} ($h_L$ and $e$), including derivation of matching formulae that take this mixing into account.
However, the full application of these formulae requires the computation of appropriate $qgq$ matrix elements, which presents a significant challenge for the lattice and hence, can be viewed as an important future direction.

\subsection{Progress in matching and renormalization}
Apart from lattice-specific systematics, which were investigated in several of the works reported above, robust extraction of partonic distributions requires addressing the issues of proper non-perturbative renormalization and reliable perturbative matching into light-cone quantities, with controlled truncation effects.

The strategy of non-perturbative renormalization of the relevant matrix elements was recently revisited.
The most widely used strategies so far involve variants of RI/MOM renormalization tailored for non-local operators \cite{Alexandrou:2017huk,Stewart:2017tvs,Izubuchi:2018srq} and ratio schemes \cite{Orginos:2017kos,Izubuchi:2018srq}.
However, it was argued in Ref.~\cite{Ji:2020brr} that such renormalization strategies introduce contaminating IR effects into large-$z$ renormalization factors.
While ratio schemes are expected to fully cancel the linear divergence, it was recently shown that RI/MOM-type schemes may evince a residual divergence \cite{Zhang:2020rsx}.
The effect was observed for several lattice setups and is smaller, but not entirely eliminated, with chiral fermions.
Obviously, it becomes numerically relevant only at rather fine lattice spacings, smaller than around 0.05-0.06 fm at the currently typical levels of precision.
Nevertheless, it supports the claim of Ref.~\cite{Ji:2020brr} that renormalization needs to be treated carefully in precision studies.

The proposal of Ref.~\cite{Ji:2020brr} is to employ a hybrid renormalization strategy, i.e.\ use the standard schemes only at perturbatively safe distances, $z\lesssim0.3$ fm, and renormalize intermediate and large distances in a different manner.
For the former, separate renormalization of linear and logarithmic divergences is advocated, with the coefficient of the linear part extractable e.g.\ from a dedicated static potential calculation.
At large distances, $z\gtrsim\Lambda_{\rm QCD}^{-1}$, unreliable in current lattice simulations, one can model the decay of the correlation with an exponential ansatz (for moderately large hadron boosts) or with a Regge-based algebraic ansatz (for large boosts).
Two applications of this strategy were reported in the conference for mesonic distributions (see next subsection).

Renormalization factors of all divergences present in the quasi-distribution matrix elements can also be disentangled from the matrix elements themselves. Having data at several lattice spacings, a strategy was proposed \cite{LatticePartonCollaborationLPC:2021xdx} how to eliminate all divergences and discretization errors. The procedure was shown to be viable using simulations with clover and overlap valence quarks on HISQ and domain wall sea, with 5 or 3 lattice spacings, ranging from 0.03 to 0.12 or 0.06 to 0.11 fm, respectively.

Renormalon effects in quasi- and pseudo-distributions were first considered in Ref.~\cite{Braun:2018brg} and revisited in Ref.~\cite{Liu:2020rqi}.
The main practical conclusion from these papers is the functional form of power corrections of $\mathcal{O}(\Lambda_{\rm QCD}^2/x^2P_z^2)$ and $\mathcal{O}(\Lambda_{\rm QCD}^2/(1-x)^2P_z^2)$, i.e.\ their enhancement both at small- and large-$x$.
It is worth to emphasize that the former translate to enhanced corrections at $x=\pm\xi$ in quasi-GPDs \cite{Bhattacharya:2018zxi,Bhattacharya:2019cme}.

On the matching side, calculations beyond NLO involving non-local operators used for quasi- and pseudo-distributions appeared recently \cite{Braun:2020ymy,Chen:2020arf,Chen:2020iqi,Chen:2020ody,Li:2020xml}.
In particular, 2-loop matching formulae were given for matrix elements renormalized in a vacuum scheme \cite{Li:2020xml}, as well as the RI/MOM and modified $\MSb$ schemes \cite{Chen:2020ody}, with estimates of numerical effects from including the NNLO terms.

The issue of origin and resummation of threshold logarithms arising in the matching was considered in Ref.~\cite{Gao:2021hxl}.
Such logarithms appear due to soft divergences in the hard coefficient functions and should be subject to all-order resummation.
However, revisiting the current pion valence PDF, the authors concluded that the effects are marginal at the current precision level.

The OPE-based matching in coordinate space was considered in Ref.~\cite{Karthik:2021sbj}.
The authors proposed an alternative procedure for such matching, i.e.\ instead of using the perturbatively computed Wilson coefficients together with their intrinsic truncation uncertainties, one can estimate them numerically based on lattice-computed matrix elements and prior knowledge of the valence PDF in the hadron.
Thus determined Wilson coefficients can then be applied to other hadrons or to other distributions in the same hadron.
The feasibility of this strategy was demonstrated using mock data and actual pseudo-PDF data for the pion and the nucleon.

\subsection{Meson distributions}
Most of the progress reported in this review concerns nucleon's partonic distributions.
However, there is considerable interest also in the structure of other hadrons, in particular mesons.
In this subsection, we briefly summarize work towards determining the $x$-dependence of mesonic distributions.

Obviously, the most interesting meson from this point of view is the pion, given its special role in QCD as the lightest particle and a pseudo-Goldstone boson.
Of particular interest is the issue of the falloff of pion PDFs at large $x$, with the postulated form of $(1-x)^\beta$.
The value of the falloff coefficient $\beta$ has been subject to much debate in the community, with various phenomenological approaches suggesting typically values close to 1 (e.g.\ JAM18 \cite{Barry:2018ort} and xFitter \cite{Novikov:2020snp} global analyses) or 2 (e.g.\ analyses based on Dyson-Schwinger equations \cite{Chen:2016sno,Bednar:2018mtf}).
However, it has been argued that the determination of $\beta$ is actually an ill-posed problem \cite{Courtoy:2020fex}, as any chosen polynomial form introduces uncontrolled dependence of $\beta$ on this form due to the feature of ``functional mimicry.''\footnote{We thank A.~Courtoy and P.~Nadolsky for their communication pointing out this issue.}
Moreover, the recent JAM global analysis with threshold resummation \cite{Barry:2021osv} found that different treatment of resummation can lead to $\beta$ values between 1 and significantly above 2.
Thus, we choose not to quote any values of $\beta$ from the recent lattice studies.

Lattice calculations of the $x$-dependence of pion PDFs were attempted with several approaches and their status has been extensively summarized in the review of Ref.~\cite{Constantinou:2020pek}.
Hence, we only mention here the most recent work presented at the conference.
The comprehensive analysis of Ref.~\cite{Gao:2020ito} for the valence pion PDF was significantly extended \cite{Gao:LAT21} by going to the physical pion mass with both non-chiral and chiral fermions, as well as testing the effects of a 2-loop matching.
The reported results indicate rather mild systematic effects from the considered sources, i.e.\ the pion mass, discretization errors ($a\in[0.04,0.076]$ fm for clover on HISQ, $a=0.11,\,0.19$ fm for domain wall), effects of using a chiral fermion and NNLO matching, leading to good agreement between the final extracted PDF and recent phenomenological determinations.
The same group presented also their preliminary results for the pion PDF with the hybrid renormalization and NNLO matching \cite{Zhao:LAT21}, in a clover on HISQ setup with a 300 MeV pion and a fine lattice spacing of 0.04 fm.
This computation highlights that at the currently attainable levels of precision, the effects of renormalization (ratio vs.\ hybrid) can already be significantly larger than statistical errors, both in the intermediate-$z$ and in the large-$z$ regime. 
Somewhat smaller effects, comparable to statistical errors, were reported also for the NLO vs.\ NNLO matching.
Finally, a practical criterion of reliability of different $x$ regimes was proposed, based on the expected size of HTE and the statistical precision, with a preliminary estimate of $x\in[0.12,0.8]$ for the reported calculation.
For most of this $x$-range, the lattice determination agrees well with phenomenological extractions, suggesting again rather small remaining systematic effects.

The authors of Ref.~\cite{Alexandrou:2021mmi} asked the question whether the $x$-dependence can be reliably reconstructed from the accessible low moments, given the currently attainable uncertainties. They computed the 3 lowest moments (that do not suffer from mixing with lower-dimensional operators) and performed 2- and 3-parameter fits to determine the pion PDF.
The implied $x$-dependence was also tested by adding the fourth moment from model/phenomenology, the effect of which was found to be negligible.
Moreover, as a consistency check, they showed that the JAM18 PDF \cite{Barry:2018ort} reconstructed from the lowest 3 moments is consistent with the full one, although with increased uncertainty due to the missing moments.
All this points out to the conclusion that 3 moments are indeed enough at the current level of precision.
Their final PDF is qualitatively consistent with phenomenological extractions and the remaining differences are likely to be associated with the missing lattice systematics (same lattice setup as reported e.g.\ in Fig.~\ref{fig:twist3}).

Pion PDFs were investigated also in (2+1)-dimensional QCD$_2$ with $N_f=0,\,2,\,4$ and $8$ quark flavors \cite{Karthik:2021qwz}, using short-distance factorization and fitting ansatz reconstruction to get from lattice-extracted matrix elements to PDFs. This led to interesting insights about the influence of the long-distance vacuum structure on the pion valence structure, since the $N_f\leq4$ theories exhibit spontaneous global flavor symmetry breaking, while the $N_f=8$ one is IR-conformal.

Apart from PDFs, there is also considerable interest in mesonic distribution amplitudes (DAs), functions relevant for several exclusive decays, representing momentum distributions of quarks in the leading Fock state of the meson's wave function.
An example are the $B\rightarrow K^*$ and $B_s\rightarrow\phi$ decays with recently reported tensions with the Standard Model by LHCb.
A first-principle determination of DAs entering the factorization of these processes could shed considerable light on this issue.
It was recently attempted within LaMET by the LPC collaboration \cite{Hua:2020gnw} in a physical pion mass setup with clover valence on HISQ sea quarks. Renormalization was performed in the hybrid scheme and the infinite boost and continuum limits were taken from 3 boosts (1.29 to 2.15 GeV) and 3 lattice spacings (0.06 to 0.12 fm).
Results for both the longitudinally and transversely polarized cases were obtained, with the former consistent with the asymptotic form and the latter evincing considerable deviations from it.

Determination of the pion DA is an aim for calculations with the revived auxiliary heavy quark approach of Detmold and Lin \cite{Detmold:2005gg}, recently dubbed HOPE (heavy OPE).
The method is based on a modified OPE of a hadronic tensor of flavor-changing axial currents with heavy quark mass suppressing HTE.
In their recent paper \cite{Detmold:2021uru}, further theoretical background was provided, in particular the relevant Wilson coefficients for determining PDFs and DAs.
The numerical work involves at the present stage detailed results for the second \cite{Detmold:2021qln} moment of the pion DA, at 4 lattice spacings (0.04 to 0.08 fm) in the quenched approximation, with clover valence fermions corresponding to a 550 MeV pion.
Several sources of systematics were considered and the final result, in agreement with other studies, was reported with $\mathcal{O}(10\,\%)$ precision and unknown, but also $\mathcal{O}(10\,\%)$, uncertainty from quenching.
Preliminary results for the fourth \cite{Perry:LAT21} moment at one lattice spacing were also reported and are in the ballpark of earlier model calculations.

New results for the pion DA were also presented with the quasi-distribution approach \cite{Juliano:LAT21}. The authors used the clover on HISQ setup with 4 lattice spacings (0.06-0.15 fm) at two pion masses (220 and 310 MeV) to perform a combined extrapolation to the continuum and the physical pion mass.
A similar setup with a single lattice spacing (0.076 fm), but a physical pion mass, was used in a study employing fits to conformal OPE, yielding results for the pion and kaon DAs \cite{Scior:LAT21}.

\subsection{Transverse-momentum-dependent PDFs (TMDs)}
To fully characterize the 3D nucleon structure, the knowledge incorporated in GPDs needs to be complemented with information on the transverse momentum dependence of partonic distributions.
This information is embodied in TMDs, objects whose lattice exploration is also in progress on the lattice.
The crucial new aspect in TMDs with respect to PDFs or GPDs is the presence of an additional type of divergences, originating from gluon radiation.
These so-called rapidity divergences need an extra regulator and can be incorporated into a soft function, which is non-perturbative for small transverse momenta.
The soft function has an intrinsic, rapidity-independent part and a rapidity-dependent part defining the Collins-Soper (CS) kernel, governing its evolution in rapidity.
The evolution in renormalization scale, in turn, is governed by cusp and hard anomalous dimensions.

Recently, a way to access the soft function from LaMET was proposed \cite{Ji:2019sxk,Ji:2019ewn}, based on a calculation of a pseudoscalar meson form factor.
Such a form factor can be factorized into the desired intrinsic soft function and a quasi-TMD wave function (quasi-TMDWF), essentially a DA with a staple-shaped link.
This strategy was followed on the lattice by two groups.

\begin{figure}[h!]
\begin{center} 
\includegraphics[scale=0.235, angle=0]{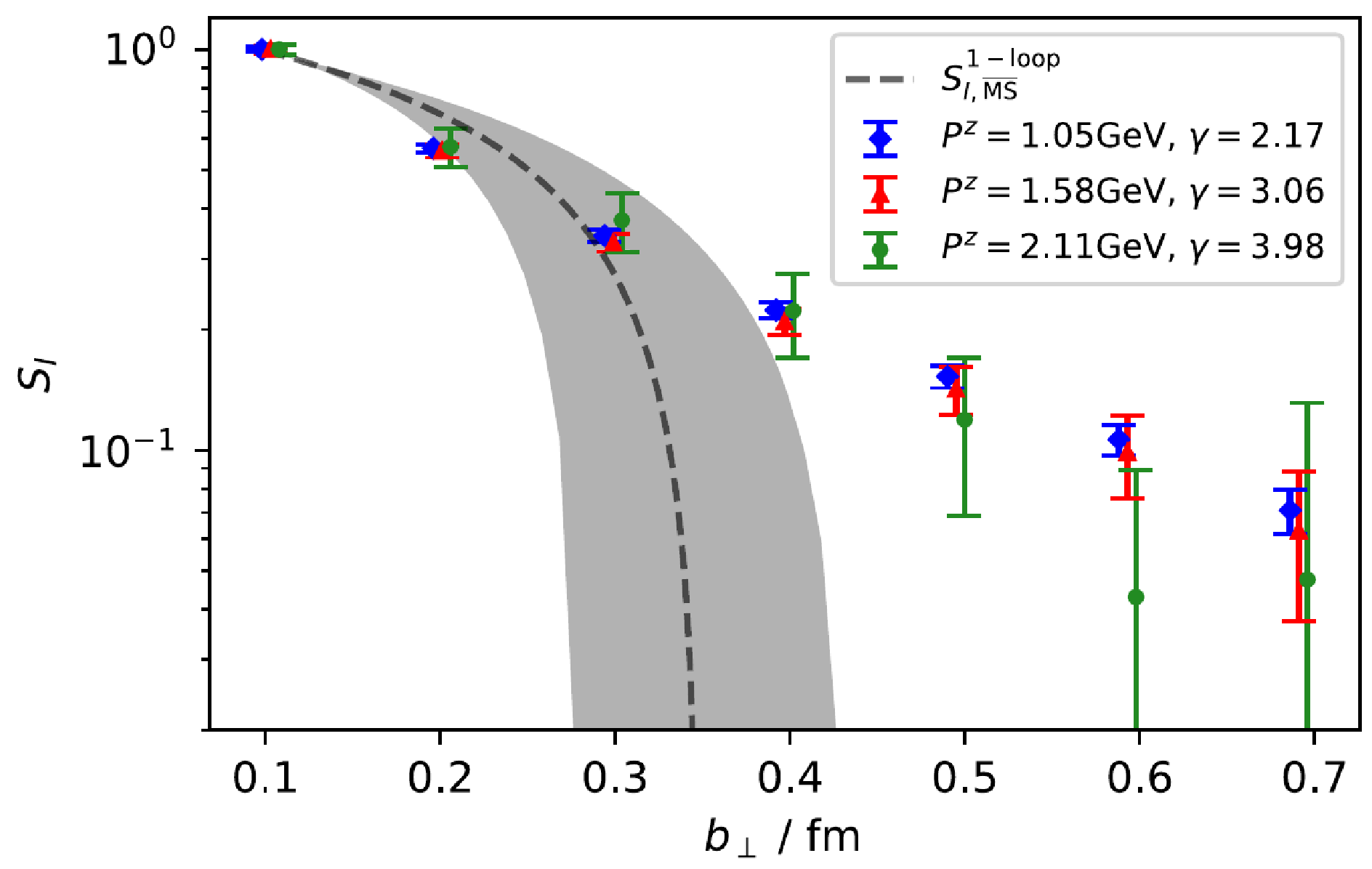}
\hspace*{-1mm}
\includegraphics[scale=0.395, angle=0]{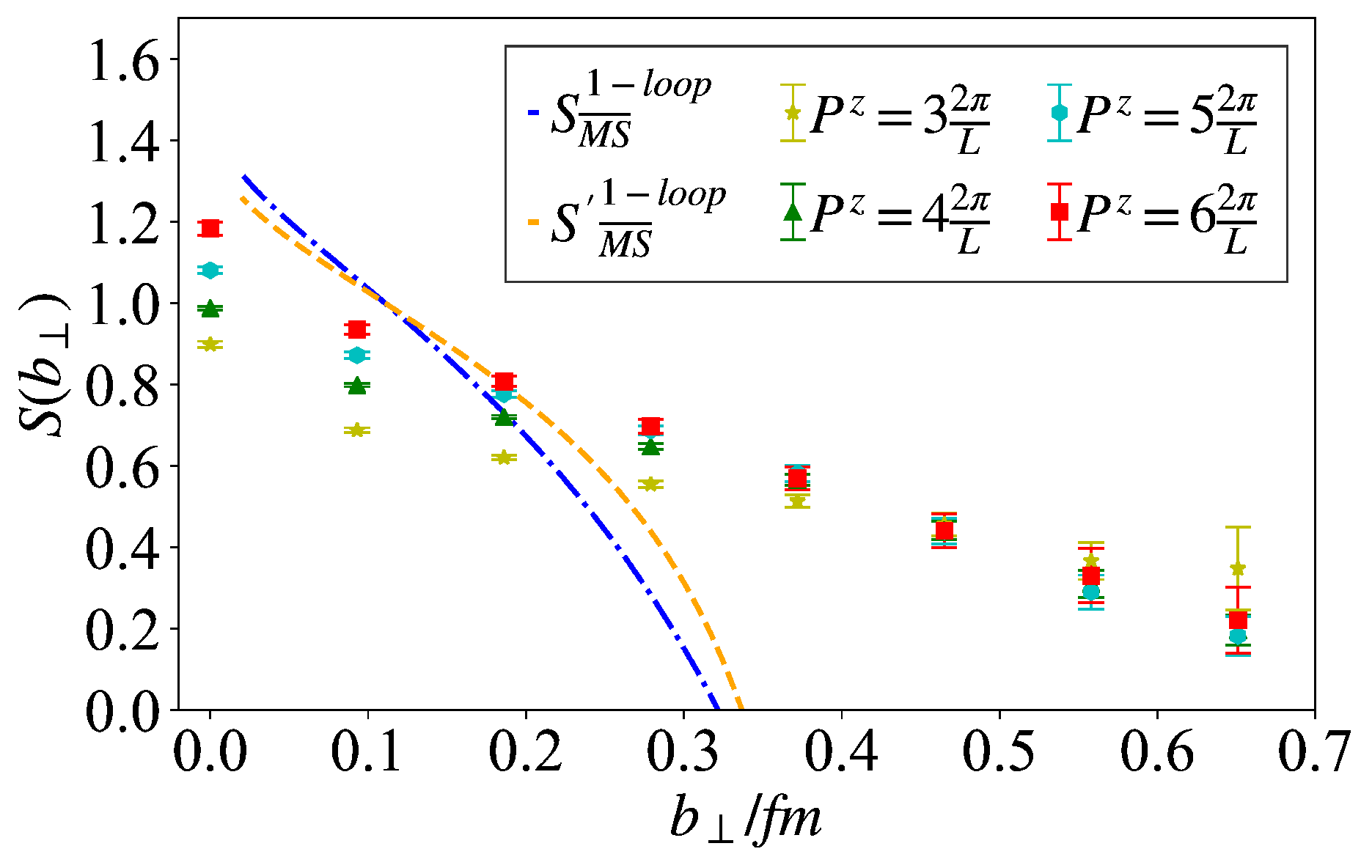}
\setlength{\tabcolsep}{3pt}
\rput(-3.5cm,5.5cm){\scriptsize\begin{tabular}{|c|c|c|c|c|c|}\hline \multirow{2}{*}{$\downarrow$} & \multirow{2}{*}{\cite{Zhang:2020dbb}} & \multirow{2}{*}{QUASI} & \multirow{2}{*}{clover} & $m_\pi^{\rm sea}\!=\!333$ MeV & \multirow{2}{*}{$a\!=\!0.098$ fm}\\&&&&$m_\pi^{\rm val}\!=\!547$ MeV&\\\hline\end{tabular}}
\rput(3.9cm,5.5cm){\scriptsize\begin{tabular}{|c|c|c|c|c|c|}\hline \multirow{2}{*}{$\downarrow$} & \multirow{2}{*}{\cite{Li:2021wvl}} & \multirow{2}{*}{QUASI} & \multirow{2}{*}{TMF} & $m_\pi^{\rm sea}\!=\!350$ MeV & \multirow{2}{*}{$a\!=\!0.093$ fm}\\&&&&$m_\pi^{\rm val}\!=\!350$-$827$ MeV&\\\hline\end{tabular}}
\end{center}
\vspace*{-1cm}
\caption{The intrinsic soft function extracted on the lattice together with 1-loop perturbative prediction. Left: LPC calculation \cite{Zhang:2020dbb}, both the lattice data and the perturbative curve are normalized to 1 at $b_\perp=a$. Right: PKU+ETMC calculation \cite{Li:2021wvl}, the two perturbative curves correspond to either 1- ($S$) or 4-loop ($S'$) value of $\alpha_s$.}
\vspace*{-3mm}
\label{fig:soft}
\end{figure}

The LPC collaboration performed the first exploratory study \cite{Zhang:2020dbb}, using a partially quenched clover setup ($a=0.098$ fm) with the valence pion heavier (547 MeV) than the sea one (333 MeV) to better isolate the signal.
They took the leading-twist Dirac structure and renormalized the soft function by forming its ratio with the same function evaluated at a fixed transverse separation of quarks ($b_\perp$), taken to be one lattice spacing.
At the present stage, only tree-level matching is known and was applied to the lattice data.
With such strategy, the soft function at $b_\perp=a$ equals its tree-level perturbative value for any nucleon boost, but its dependence on the transverse separation can be compared to the perturbative prediction at $b_\perp>a$.
The authors concluded agreement with the latter for distances up to around 0.3 fm, i.e.\ in the perturbative regime (see the left panel of Fig.~\ref{fig:soft}).
The study, thus, proved the feasibility of the method, but with its level of complexity, several systematic effects need to be explored.

\hspace*{6mm}

Some of these systematics were investigated in the second study, by a collaboration of Beijing University (PKU) with ETMC \cite{Li:2021wvl}, employing TMF, also with a single lattice spacing ($a=0.093$ fm) and a range of valence pion masses from 827 MeV down to the unitary one of 350 MeV.
The authors observed large higher-twist contamination for different Dirac structures and, using Fierz identities, formed their combinations which reduce these effects significantly.
The renormalization was done with a ratio involving an operator with the same transverse separation in both the numerator (boosted) and the denominator (at rest), allowing for a clean cancellation of divergences.
This also avoids imposing any specific value of the renormalized soft function at some $b_\perp$, thus retaining predictive power at all distances.
The right panel of Fig.~\ref{fig:soft} presents a comparison with the perturbative curve.
Large dependence on the nucleon boost was revealed for small $b_\perp$ and upon extrapolation to infinite momentum, approximate agreement with the 1-loop curve could be concluded.
Qualitative agreement was evinced also with the LPC calculation, with quantitative one hindered by the different renormalization procedures.
Ref.~\cite{Li:2021wvl} also confirmed the theoretically conjectured independence of the results on the pion mass.
Despite addressing some systematic effects, several more need to be scrutinized, notably effects of matching beyond tree-level.

\begin{figure}[h!]
\begin{center} 
\includegraphics[scale=0.4, angle=0]{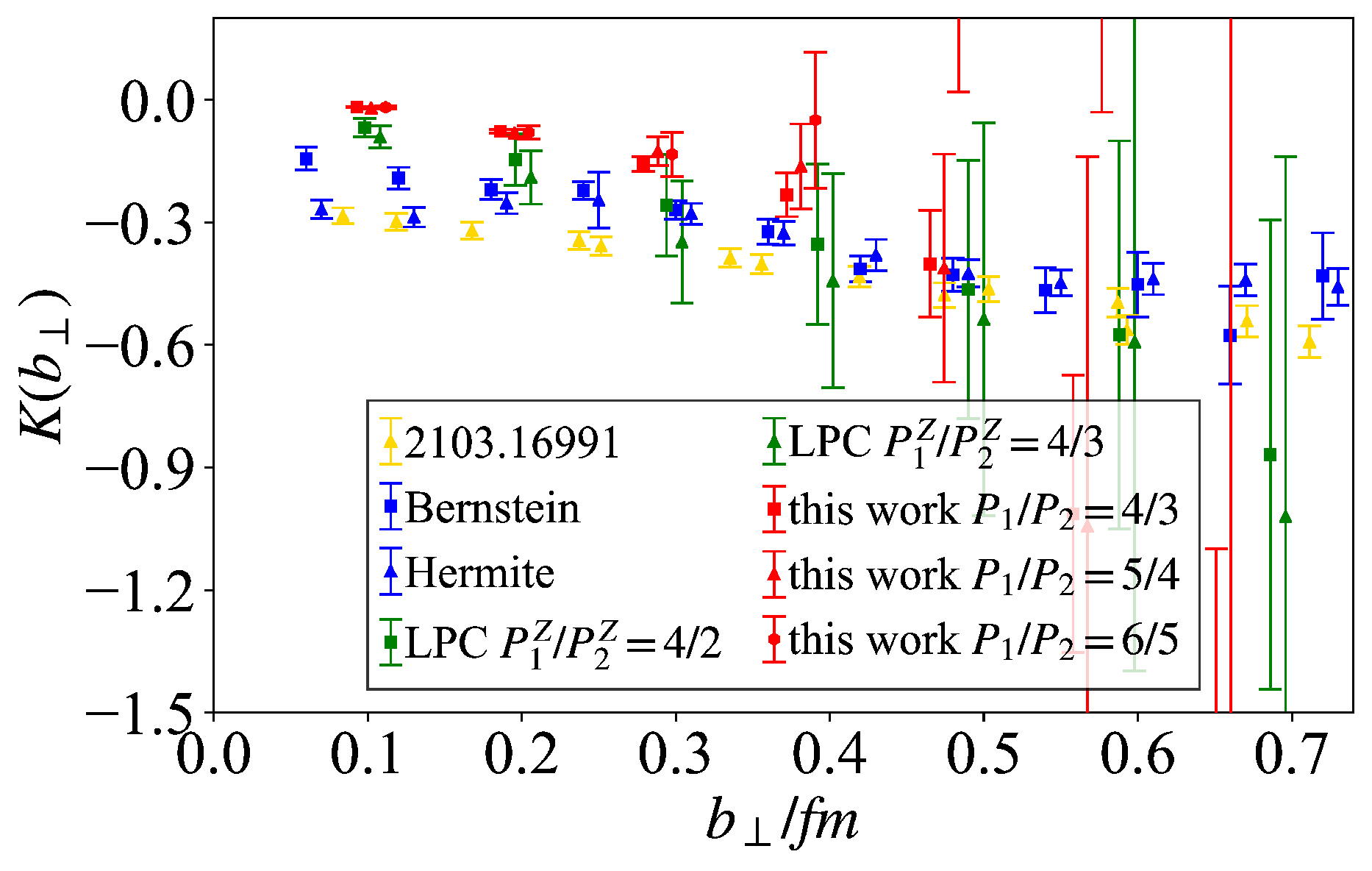}
\hspace*{-1mm}
\includegraphics[scale=0.8, angle=0]{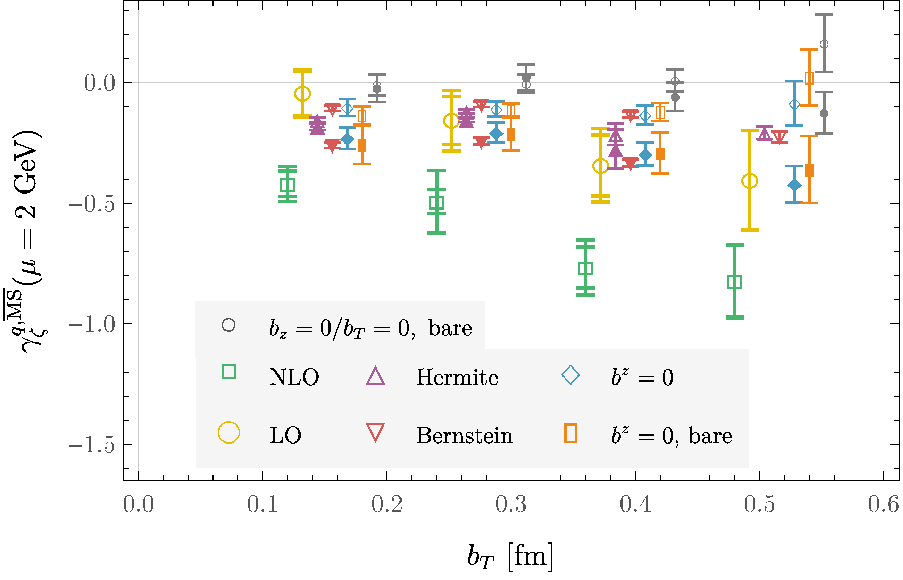}
\setlength{\tabcolsep}{3pt}
\rput(-3.6cm,5.87cm){\scriptsize\begin{tabular}{|c|c|c|c|c|c|}\hline \multirow{4}{*}{$\downarrow$} & \cite{Zhang:2020dbb} & \multirow{3}{*}{QUASI} & clover & $m_\pi^{\rm val}\!=\!547$ MeV & $a\!=\!0.098$ fm\\\cline{2-2}\cline{4-6}
 & \cite{Li:2021wvl} & & TMF & $m_\pi^{\rm val}\!=\!350$-$827$ MeV & $a\!=\!0.093$ fm\\\cline{2-2}\cline{4-6}
 & \cite{Shanahan:2020zxr} & & quench. & $m_\pi^{\rm clover}\!=\!1.2$ GeV & $a\!=\!0.06$ fm\\\cline{2-6}
 & \cite{Schlemmer:2021aij} & MELLIN & clover & $m_\pi\!=\!422$ MeV & $a\!=\!0.085$ fm\\\hline\end{tabular}}
\rput(3.95cm,5.43cm){\scriptsize\begin{tabular}{|c|c|c|c|c|c|}\hline $\downarrow$ & \cite{Shanahan:2021tst} & QUASI & clover on HISQ&$m_\pi^{\rm val}=538$ MeV & $a\!=\!0.12$ fm\\\hline\end{tabular}} 
\end{center}
\vspace*{-1cm}
\caption{The Collins-Soper kernel. Left: Determinations from Refs.~\cite{Schlemmer:2021aij} (``2103.16991''), \cite{Shanahan:2020zxr} (``Hermite'', ``Bernstein''), \cite{Zhang:2020dbb} (LPC) and \cite{Li:2021wvl} (``this work''). Right: Comparison of several methods employing the same lattice data \cite{Shanahan:2021tst}. For details, we refer to the original papers.}
\vspace*{-3mm}
\label{fig:CS}
\end{figure}

The rapidity-dependent part of the soft function, i.e.\ the CS kernel, was subject to several recent determinations.
In particular, as proposed in Ref.~\cite{Ebert:2018gzl}, it can be accessed with a ratio of quasi-TMDs at different rapidities. 
This method was first explored numerically \cite{Shanahan:2020zxr} in the quenched approximation with clover valence quarks ($a=0.06$ fm, $m_\pi^{\rm val}=1.2$ GeV).
A more advanced study was presented in Ref.~\cite{Shanahan:2021tst}, in a clover on HISQ setup ($a=0.12$ fm, $m_\pi^{\rm val}=538$ MeV).
The authors compared several analysis methods using the same lattice data and concluded an important role of power corrections and NLO perturbative matching that was applied for the first time.
The above mentioned ratio of quasi-TMDs can also be rewritten as a ratio of quasi-TMDWFs at different boosts, which need to be computed for the determination of the intrinsic soft function.
Thus, the CS kernel could naturally be accessed also in Refs.~\cite{Zhang:2020dbb,Li:2021wvl}.
Preliminary results from a follow-up LPC study of the CS kernel were also presented in the conference \cite{Chu:LAT21}.
An alternative approach is to calculate the CS kernel from ratios of first Mellin moments of TMDs \cite{Schlemmer:2021aij}, numerically demonstrated with clover fermions ($a=0.085$ fm, $m_\pi=422$ MeV).
The results from all these studies, except for Ref.~\cite{Shanahan:2021tst}, are compared in the left panel of Fig.~\ref{fig:CS}.
The overall qualitative agreement between these results is encouraging (in particular between the two studies aiming at the intrinsic soft function that both utilize quasi-TMDWFs), but it is clear that any quantitative conclusions need to be postponed until errors are properly estimated and extrapolated out.
The right panel emphasizes this conclusion and suggests important role of non-lattice systematics -- notably of the matching (NLO vs.\ LO), as hinted above.\vspace*{11mm}

Nevertheless, all the above studies convincingly show the feasibility and progress of TMDs calculations on the lattice. 
We also mention the related theoretical LaMET studies of single transverse-spin asymmetry and the Sivers function \cite{Ji:2020jeb} and of light-front wave functions \cite{Ji:2021znw}.

\subsection{Other developments}
The recent work on $x$-dependent hadron structure was not limited to the above reported papers. Here, we briefly mention some other recent developments.

As thoroughly discussed in Ref.~\cite{Karpie:2019eiq}, the issue of reconstruction of the $x$-dependence from lattice data suffers from the inverse problem, originating from the fact that a continuous distribution is to be determined from a finite set of truncated lattice data.
In addition to methods of overcoming this issue suggested in this paper, another possibility of employing the Bayes-Gauss-Fourier transform was proposed in Ref.~\cite{Alexandrou:2020tqq} and tested on the physical point ETMC lattice data of Refs.~\cite{Alexandrou:2018pbm,Alexandrou:2019lfo}.

FVE in non-local matrix elements were considered initially a few years ago \cite{Briceno:2018lfj}, where it was shown in a toy scalar model that the non-local operator can lead to replacement of the control parameter of $m_\pi L$ by $m_\pi(L-z)$ and thus, to an enhancement of FVE.
In a follow-up paper \cite{Briceno:2021jlb}, the authors presented a model-independent framework to determine FVE of current-current operators, employing lattice data on relevant elastic form factors.
The procedure was illustrated in 2- and 4-dimensional scalar theories.

Another insight on FVE came in the framework of chiral perturbation theory \cite{Liu:2020krc}, concluding rather complicated dependence of FVE on the volume, but its strong suppression with the nucleon boost.
FVE are always smaller for a boosted nucleon compared to the rest frame and below 1\% as soon as $m_\pi L\gtrsim3$.
The paper provided also formulae for the pion mass dependence of matrix elements underlying quasi- and pseudo-PDFs, which can be used to guide the extrapolation from non-physical pion masses to the physical one.

The final paper that we mention \cite{DelDebbio:2020cbz} is a study of renormalization and factorization of Euclidean matrix elements underlying quasi- and pseudo-PDFs, performed in a nongauge theory (6-dimensional $\phi^3$ theory). This allowed the authors to highlight the key issues without the complications of QCD and to establish a formal equivalence between factorizations in coordinate and momentum space. Alternative observables subjectable to factorization were also defined using gradient flow. 
The multitude of formally equivalent (but yet with different lattice and other systematics) Euclidean observables which make access to light-cone PDFs possible upon factorization suggests the feasibility of the program already suggested earlier by Ma and Qiu \cite{Ma:2014jla,Ma:2017pxb}, of a global analysis of all lattice data, in analogy to global fits of experimental cross sections.

\section{Summary and prospects}
\label{sec:summary}
Lattice QCD has been providing insights into the structure of hadrons for many years.
However, the field of $x$-dependent partonic distributions is relatively new, with intensive studies started only around 8 years ago.
During this time, enormous progress has been achieved with a plethora of theoretical and practical approaches.
There is an increasing number of distributions accessible on the lattice.
For some of them, only exploratory studies are available, but for some other, already several systematic uncertainties are being investigated, leading to the prospect of precision calculations in the near future. In this final section of this review, we offer some remarks about these prospects and the directions for further work. We group them in 3 categories.\vspace*{2mm}

\noindent\textbf{Robustness and reliability of the lattice extraction}.
There are several requirements for a robust extraction of partonic distributions on the lattice.
Obviously, one of the most important ones is the statistical quality of the signal.
This presents arguably the biggest challenge for the future -- how to reliably reach large nucleon boosts, necessary for safe contact between Euclidean observables and light-cone distributions, under the generic problem of exponential decrease of the signal with increasing boost.
The problem is aggravated when simulating at the physical pion mass, which is due to increasing excited states contamination.
The latter necessitates working at large source-sink separations, implying further exponential worsening of the signal.
Major progress has been achieved via the application of the momentum smearing technique, but it is important to understand that it does not cure the exponentially hard problem, but only moves it to larger boosts.
In the universal experience of almost all groups (see Section \ref{sec:results}), small statistical errors for highly-boosted hadrons at the physical pion mass are still out of reach.
Thus, an improvement in lattice techniques is highly desired.
An attempt to overcome this problem is the distillation program of the HadStruc collaboration, offering better control over excited states and leading in practice to the possibility of extracting the  matrix elements of interest with smaller source-sink separations.
However, it is yet to be demonstrated how high in momentum one can go at the physical point.
Alternative/complementary methods are, clearly, desirable.
It should also be kept in mind that the boost should not be larger than the lattice cutoff -- thus, larger boosts will require also finer lattice spacings and larger lattices.

Another aspect, rather general for all lattice computations, is reliable control over all sources of lattice systematics, such as discretization effects or FVE.
For the former, it is crucial to implement an $\mathcal{O}(a)$-improvement program.
While first continuum limit studies begin to appear, the non-locality of the employed operators induces $\mathcal{O}(a)$ effects, effectively leading to large uncertainties in the continuum limit.
Furthermore, broader systematics of the lattice extractions are also important.
Recent important progress concerns non-perturbative renormalization and truncation effects in the matching.
However, other vital issues remain -- in particular, it is desirable to control better the higher-twist contamination.
The ``simple'' way is to work at large hadron boosts, naturally suppressing the power corrections.
This approach rather quickly hits a wall of very bad signal quality with the current lattice techniques.
Alternatively, one could explicitly calculate these HTE and subtract them, which is a potentially interesting direction, requiring the formulation of a viable implementation and derivation of appropriate matching formulae.
Yet different important issue is the reconstruction of the $x$-dependence given the fact that a continuous distribution needs to be determined from a finite number of calculated inputs and moreover, truncated at some separation.
This inverse problem seems not too severe with the current uncertainties, but may become more pressing when errors are reduced.
Several methods of alleviating the inverse problem have been proposed and combined with a prospective better quality of data, one can expect robust reconstruction.

In general, one needs to observe the hierarchy of errors and address first the ones that lead to largest uncertainties, while at some level of precision, treatment of the subleading ones will become an issue.\vspace*{2mm}

\noindent\textbf{Exploration of new directions}.
As already argued, an increasing number of distributions is becoming accessible on the lattice.
The recent 1-2 years have brought first investigations of hitherto unexplored cases, such as GPDs, twist-3 PDFs or the soft function.
It is expected that these calculations will soon enter a more mature stage with quantified systematics.
Additionally, partonic distributions in other hadrons can be pursued.
So far, most analyses focused on the phenomenologically most relevant cases of the nucleon and to a lesser extent the pion and the kaon.
However, other hadrons are also of phenomenological interest, such as the above reported $K^*$ and $\phi$ mesons and with other ones that can follow.
Other baryons may also be of interest, such as the $\Delta^+$ \cite{Chai:2020nxw} that can shed light on the sea quark asymmetry in the nucleon.\vspace*{2mm}

\noindent\textbf{Synergy of lattice and phenomenology}.
Ultimately, it is natural to expect some impact of this rich lattice program on phenomenology.
The prerequisite for this is that the reconstructed distributions have properly estimated all sources of systematics.
At present, this condition is not satisfied in any lattice determination, but one can expect that this may change rather soon.
Obviously, it is desired that the overall uncertainty is also as small as possible.
With this in mind, it is unlikely that lattice can have significant impact for unpolarized twist-2 PDFs, the case with huge abundance of experimental data and experience in global fits.
Thus, these PDFs are expected to be the benchmark case, with the lattice striving to reproduce the PDFs from global fits with fully quantified and decreasing total uncertainties.
However, for the less known distributions, lattice inputs with even $\mathcal{O}(20\%)$ total error may have an impact.
In fact, this concerns several cases, like transversity PDFs and GPDs/TMDs or higher-twist PDFs in general.
One should keep in mind that their experimental status shall change as well with the advent of new experimental facilities, such as the EIC.
This opens up the possibility of a truly complementary role of lattice and phenomenology.

\section*{Acknowledgments}
I would like to thank the Organizers of LATTICE2021 for the invitation to present this review talk and for the very successful and enjoyable conference.
I also thank M.~Constantinou, R.~Sufian, Y.-B.~Yang, S.~Zafeiropoulos and Y.~Zhao for discussions/material covered in my talk.
Special thanks to M.~Constantinou for carefully reading the manuscript and offering important and useful suggestions.
I am grateful also to all of my Collaborators in projects whose results are reported in this review: C.~Alexandrou, M.~Bhat, S.~Bhattacharya, Y.~Chai, M.~Constantinou, L.~Del Debbio, J.~Dodson, X.~Feng, T.~Giani, J.~Green, K.~Hadjiyiannakou, K.~Jansen, G.~Koutsou, Y.~Li, Ch.~Liu, F.~Manigrasso, A.~Metz, A.~Scapellato, F.~Steffens, J.~Tarello, S.-C.~Xia.
Financial support from the National Science Centre (Poland) is acknowledged, grant SONATA BIS no.\ 2016/22/E/ST2/00013.

\bibliographystyle{h-physrev}
\bibliography{references}

\end{document}